\documentclass[a4paper,twocolumn, aps, pra]{revtex4-2}

\pdfoutput=1
\usepackage[utf8]{inputenc}
\usepackage[english]{babel}
\usepackage[T1]{fontenc}
\usepackage{amsmath}
\usepackage{hyperref}
\usepackage{orcidlink}

\usepackage{graphicx}
\usepackage{url}
\usepackage{float}
\usepackage{amsmath}
\usepackage{amsfonts}
\usepackage{physics}
\usepackage{dsfont}
\usepackage{bm}
\usepackage{hyperref}
\usepackage[capitalise]{cleveref}

\usepackage{tikz}
\usetikzlibrary{quantikz2}

\hypersetup{
  linkcolor  = blue,
  citecolor  = red,
  urlcolor   = magenta,
  colorlinks = true,
}

\usepackage{changes}

\definechangesauthor[name=jannes, color=red]{jannes}

\definechangesauthor[name=stef, color=green]{stef}

\definechangesauthor[name=rod, color=blue]{rod}

\let\vec\bm

\newcommand{\iden}{\mathbb{I}}
\newcommand{\G}[2]{\Gamma^{#1}_{#2}}
\newcommand{\tG}[1]{\tilde{\Gamma}_{#1}}
\newcommand{\eqconstr}{\overset{c}{=}}

\usepackage{tocbasic}
\DeclareTOCStyleEntry[
  beforeskip=.5em plus 1pt,% default is 1em plus 1pt
  pagenumberformat=\textbf
]{tocline}{section}

\begin{document}

\title{Local fermion-to-qudit mappings: a practical recipe for four-level systems}

\author{Rodolfo Carobene \orcidlink{0000-0002-0579-3017}}
\affiliation{Department of Physics, University of Milano-Bicocca, I-20126 Milano, Italy}
\affiliation{INFN - Milano Bicocca, I-20126 Milano, Italy}
\affiliation{Bicocca Quantum Technologies (BiQuTe), I-20126 Milano, Italy}

\author{Stefano Barison \orcidlink{0000-0001-5842-1113}}
\affiliation{Institute of Physics, École Polytechnique Fédérale de Lausanne (EPFL), CH-1015 Lausanne, Switzerland}
\affiliation{%
Center for Quantum Science and Engineering, \'{E}cole Polytechnique F\'{e}d\'{e}rale de Lausanne (EPFL), CH-1015 Lausanne, Switzerland
}
\affiliation{National Centre for Computational Design and Discovery of Novel Materials MARVEL, EPFL, Lausanne, Switzerland}

\author{Andrea Giachero \orcidlink{0000-0003-0493-695X}}
\affiliation{Department of Physics, University of Milano-Bicocca, I-20126 Milano, Italy}
\affiliation{INFN - Milano Bicocca, I-20126 Milano, Italy}
\affiliation{Bicocca Quantum Technologies (BiQuTe), I-20126 Milano, Italy}

\author{Jannes Nys \orcidlink{0000-0001-7491-3660}}
\affiliation{Institute for Theoretical Physics, ETH Zurich, 8093 Zurich, Switzerland}

\begin{abstract}
    In this paper, we present a new set of local fermion-to-qudit mappings for simulating fermionic lattice systems. We focus on the use of multi-level qudits, specifically ququarts. Traditional mappings, such as the Jordan-Wigner transformation (JWT), while useful, often result in non-local operators that scale unfavorably with system size. To address these challenges, we introduce mappings that efficiently localize fermionic operators on qudits, reducing the non-locality and operator weights associated with JWT. We propose one mapping for spinless fermions and two mappings for spinful fermions, comparing their performance in terms of qudit-weight, circuit depth, and gate complexity. 
    Notably, for superconducting devices, the control complexity of qudits compared to qubits does not increase prohibitively for four-level systems, making them practical for near-term demonstrations, though control challenges grow with increasing dimensionality.
    Therefore, we provide solutions to the operator decompositions of the Trotterized quantum dynamics into one and two-qudit gates for all mappings.
    By leveraging the extended local Hilbert space of qudits, we show that these mappings enable more efficient quantum simulations in terms of two-qudit gates, reducing hardware requirements without increasing computational complexity. We validate our approach by simulating prototypical models such as the spinless $t$-$V$ model and the Fermi-Hubbard model in two dimensions, using Trotterized time evolution. Finally, we show the connection between the plaquette constraint of our mapping and the $\mathbb Z_ 2$ Toric code. This connection can be exploited to prove that the ground state of the localized qudit mapping can be efficiently prepared using quantum circuits or measurement-based feedback control. Our results highlight the potential of qudit-based quantum simulations in achieving scalability and efficiency for fermionic systems on near-term quantum devices.
\end{abstract}

\maketitle

%\setcounter{secnumdepth}{1}
%\setcounter{tocdepth}{1}
%\tableofcontents
\section{Introduction}
Quantum simulation of fermionic systems represents one of the most promising applications of digital quantum devices.
However, they present significant challenges due to the anti-commutation relations of fermionic operators. 
Specifically, to conduct such simulations, one must map anti-commuting fermion operators onto Pauli operators $\{\sigma^x, \sigma^y, \sigma^z\}$, which inherently obey commutation relations $[\sigma^\mu, \sigma^\nu] = 2 i \epsilon^{\mu \nu \kappa} \sigma^\kappa$.

The most commonly used mapping between fermionic and spin degrees of freedom is the Jordan-Wigner transformation (JWT), which naturally arises from the second quantization formalism of fermions~\cite{jordan1993paulische}. 
While physical fermionic Hamiltonians are typically \emph{local} and consist of even-fermion parity operators, the JWT can map these local operators onto non-local Pauli (string) operators.
This is due mainly to the introduction of transformation properties for the individual fermionic operators.
These string operators scale with the system size for $d>1$ dimensional lattice Hamiltonians, eventually rendering the calculations challenging for large systems sizes~\cite{whitfield2016local, havlivcek2017operator}. 
Recent theoretical advancements have focused on new methodologies to simulate 2D Jordan-Wigner-transformed fermionic systems on contemporary noisy devices, prominently using Fermionic-SWAP gates (FSWAP) introduced in Ref.~\cite{kivlichan2018quantum} and further elaborated in Ref.~\cite{cade2020strategies}. 
However, such methods require an increasing number of FSWAP operations as the system size or dimension grows, posing challenges for digital quantum simulations for large scale quantum simulations.

Fortunately, the JWT is not the only mapping available and there is the possibility of searching for efficient encodings that require fewer qubits and/or reduce the operator weights~\cite{derby2021compact, jiang2020optimal, bravyi2002fermionic, steudtner2018fermion, setia2019superfast}. 
Over the past 40 years, various works have focused on generalizing mappings to higher dimensions (>1D) to maintain \emph{locality} in the operators and reduce the size of the Jordan-Wigner strings. 
One of the earliest studies to derive higher-dimensional JWT generalizations is the work of Wosiek~\cite{wosiek1981local}. 
Ball~\cite{ball2005fermions} and Verstraete-Cirac~\cite{verstraete2005mapping} independently explored similar ideas, introducing auxiliary fermions to expand the fermionic Hilbert space and then defining an auxiliary Hamiltonian to restrict the reachable Hilbert space. 
Compared to Ref.~\cite{wosiek1981local}, the Ball-Verstraete-Cirac (BVC) transformations~\cite{ball2005fermions, verstraete2005mapping} explicitly introduced auxiliary fermionic modes to counteract the Jordan-Wigner strings. 
These auxiliary modes effectively store the parity near the interaction terms, otherwise captured by the Jordan-Wigner string~\cite{whitfield2016local}, resulting in local qubit Hamiltonians.
Given the potential benefits for quantum simulations, in recent years there has been a renewed interest in these localized approaches~\cite{setia2018bravyi, setia2019superfast, chen2018exact, chen2023equivalence, bochniak2020bosonization, li2021higher, po2021symmetric, derby2021compact,  derby2021compact2, o2024local, chiew2021optimal, chen2020exact, jafarizadeh2024recipe, algaba2024low,wyrzykowski2022thesis,Ogura2020,Chapman2020,Minami2016}, in particular through its interpretation as lattice gauge theories~\cite{chen2018exact, chen2023equivalence, fradkin1980fermion, ballarin2024digital, srednicki1980hidden}. 
Unlike earlier methods (such as Refs.~\cite{ball2005fermions, verstraete2005mapping}) that explicitly perform JWT, recent techniques define bosonic operators from fermionic ones, which are then mapped directly onto qubit operators without ordering the fermions. 
Since these novel methods do not require fermion ordering, they can be more easily generalized to different systems and higher dimensions. 
However, increasing the Hilbert space introduces constraints that must be fulfilled exactly, and it was pointed out in Refs.~\cite{nys2023quantum, nys2022variational} that quantum circuits that minimally fulfill these constraints scale with the system size. 
Ref.~\cite{guaita2024locality} further highlighted that while local fermion-to-qubit mappings introduce local Hamiltonians, they require a high degree of state non-locality. However, recent work in Ref.~\cite{nigmatullin2024experimental} has experimentally demonstrated the feasibility and power of local encodings over Jordan-Wigner mappings. On larger systems, initial states that satisfy the constraints (i.e.\ a toric code state) can be prepared in constant depth using measurement and
feed-forward control~\cite{iqbal2024topological, foss2023experimental}.

The use of multi-level quantum systems for quantum simulation has been gaining a lot of attention~\cite{lanyon2008quantum,chi2022,Ringbauer2022,Balantekin2024,Sawaya2020}. Recent work has explored its properties for simulating electronic systems~\cite{fischer2023universal,Vezvaee2024,Chizzini2024}.
These systems, known as \emph{qudits}, are becoming an important alternative due to a number of potential improvements with respect to qubit devices~\cite{neeley2009emulation, tacchino2021proposal, kurkcuoglu2021quantum, GonzlezCuadra2022, meth2023simulating}. 
These advantages include increased error resilience, and quantum error corrections with reduced code sizes~\cite{muralidharan2017overcoming,campbell2014enhanced}. 
It has been demonstrated in various works that the use of qudits can significantly reduce the complexity of quantum circuits by reducing the number of two-qudit gates compared to their qubit representations~\cite{fischer2023universal}. Indeed, joining multiple degrees of freedom into the local Hilbert space of a single qudit can decrease the number of inter-qudit operations. 

In addition, single-qudit gates are generally faster and exhibit higher fidelity compared to multi-qudit entangling operations~\cite{Krantz2019,Gao2021}. Specifically for superconducting ququarts, increasing the local dimension tends to increase the number of single-qudit gates required, while simultaneously reducing the total number of entangling gates needed~\cite{fischer2023universal}. From a hardware perspective, employing qudits does not significantly increase the complexity of control electronics. For instance, in superconducting quantum processors, qubits and qudits share essentially the same control hardware and number of control lines. The main additional overhead arises from longer measurement sequences needed to distinguish multiple energy levels, along with modest increases in pulse-shaping and calibration efforts to selectively address transitions between adjacent levels~\cite{Kononenko2021,Seifert2023,Liu2023,Wang2024}.

In this work, we demonstrate how to simulate higher-dimensional fermionic lattice models on a lattice of four-level qudits (ququarts), by introducing a set of local fermion-to-qudit mappings tailored specifically to ququarts. The choice of four-level systems is motivated by their ability to naturally address the problem of locality of fermionic mappings -- particularly for electron-like fermions -- by enabling a fully local encoding of fermionic parity. This is achieved by exploiting the increased dimensionality to host ancilla-like degrees of freedom within each ququart, thus avoiding non-local parity strings.

This work provides a practical and physically motivated recipe for constructing local mappings that satisfy the required (anti-)commutation relations. We focus on different ququart partitionings to encode the two fermionic spin sectors, exploring the trade-offs and advantages offered by each mapping. Our methodology lays the groundwork for a more general framework for mapping fermionic systems to multi-level qudit representations.

While previous studies have investigated qudit-based mappings for fermionic systems~\cite{Vezvaee2024,meth2023simulating,Chizzini2024} proving their feasibility, our approach introduces several strictly-local new constructions that are simple to implement both for the spinless and spinful case, and we provide efficient gate decomposition strategies suitable for Trotterized dynamics. 
We also present a comparison of resource requirements, including two-qudit gate counts, contrasting our ququart-based mappings with the non-local Jordan-Wigner transformation, state-of-the-art local fermion-to-qubit mappings and recent fermion-to-qudit mapping proposals~\cite{Vezvaee2024,Chizzini2024}.
Finally, in Sec. \ref{sec:toric_code}, we demonstrate that satisfying the plaquette constraints relates directly to preparing a Toric code ground state.

\section{Fermionic operators}\label{sec:fermionic_operators}
We will consider a set of fermionic modes with index $i$ with respective creation $f_i^\dagger$ and annihilation $f_i$ operators.
To each mode we associate the Majorana operators
\begin{align}
    \gamma_i &:= f_i^\dagger + f_i \\
    \gamma_i' &:= i(f_i^\dagger - f_i) 
\end{align} 
% or inversely
% \begin{align}
% f_i^\dagger &= \frac{1}{2} (\gamma_i - i \gamma_i') \\
% f_i &= \frac{1}{2}(\gamma_i + i \gamma_i' ) \\
% \end{align}
that fulfill the anti-commutation relations
\begin{align}
    \{\gamma_i, \gamma_j\} &= \{\gamma_i', \gamma_j'\} = 2\delta_{ij}\\
    \{\gamma_i, \gamma_j'\} &= 0
\end{align}
In order to map the fermionic system onto qudit operators, we will rewrite the relevant even-parity fermion operators and determine their commutation properties.
A hopping term between two fermionic modes $i \neq j$ becomes
\begin{equation}
\begin{split}
    f_i^\dagger f_j + f_j^\dagger f_i &= \frac{i}{2}(\gamma_i\gamma_j' + \gamma_j \gamma_i') \\
    &= \frac{1}{2} \left(S_{ij} + S_{ji}\right)
\end{split}
\end{equation}
where we introduced $S_{ij} := i\gamma_i \gamma_j'$.
The number and parity operators read
\begin{align}
    n_i &= f_i^\dagger f_i = \frac{1}{2}(1 + i \gamma_i \gamma_i') \\
    P_i &= (-1)^{n_i} = 1-2n_i = -i \gamma_i\gamma_i'
\end{align}
A pair of fermions can be added by considering the even-parity operator
\begin{align}
    f_i^\dagger f_j^\dagger &= \frac{1}{4}\left( \gamma_i \gamma_j - \gamma_i' \gamma_j' - i\gamma_i \gamma_j' -i \gamma_j \gamma_i' \right)
\end{align}

%We then introduce the edge operators:
%\begin{equation}
%    S_{ij} = i\gamma_i\gamma_j'
%\end{equation}
%so that we have the equality $f_i^\dagger f_j + f_j^\dagger f_i = S_{ij} + S_{ji}$.
%From this we can obtain the commutation relation for $S_{ij}$ ($i\neq j \neq k \neq l$):
%\begin{align}
%    [S_{ij}, S_{jk}] = [S_{ij}, S_{kl}] = 0 \\
%    \{S_{ij}, S_{ik}\} = \{S_{ij}, S_{kj}\} = 0
%\end{align}
%Putting this into words: edge operators that share the start (end) vertex anti-commute, while operators on completely different vertices or for which the start vertex is the end one for the other operator, behave bosonic and commute.

We can decompose the edge operators $S$ by introducing the edge and vertex operators
\begin{align}
    A_{ij} &:= -i \gamma_i \gamma_j \\
    B_{k} &:= -i \gamma_k \gamma_k'
\end{align}
leading to $S_{ij}=-iA_{ij}B_j$ and
\begin{align}
S_{ij} + S_{ji} &= -i A_{ij} (B_i  - B_j) .
\end{align}

% a graphical representiation would be nice here
From the above definition, we observe that $A$ is anti-symmetric in the mode indices $A_{ji} = -A_{ji}$. Furthermore, in order to respect the fermionic anti-commutation relations, it can be shown that the $A$ and $B$ operators must obey the following rules
\begin{itemize}
\item $[A_{ij},A_{kl}]=0$ if $i \neq j \neq k \neq l$,
\item $[A_{ij},B_{k}]=0$ if $i \neq j \neq k$,
\item $[B_{i},B_{j}]=0$ if $i \neq j$,
\item $\{A_{ij},A_{kl}\}=0$ if $(i,j)$ and $(k,l)$ have an overlapping mode, e.g.\ $j=k$,
\item $\{A_{ij},B_{k}\}=0$ if $i=k$ or $j=k$,
\end{itemize}
in other words: if pairs of link and/or vertex operators do not share a fermionic mode index, they \emph{commute} as bosonic operators, but they \emph{anti-commute} when they share an index.

All even fermionic operators can be expressed in terms of the edge and vertex operators, for example, we can see that the $B$ operator corresponds to the parity. We can also create particle pairs using
\begin{align}
f_i^\dagger f_j^\dagger &= \frac{i}{4} A_{ij}(1 + B_i)(1+B_j) 
\end{align}
Finally, it will later be useful to introduce the behavior of the fermionic operators on a loop of fermionic modes. Consider a loop of fermionic modes $\mathcal{P} = \{1, \dots, \abs{\mathcal{P}}\}$, and apply the identity $\gamma_i^2 = \iden$ to all fermionic modes on the loop, then after regrouping we obtain
\begin{equation}\label{eq:identity_prod_A}
\begin{split}
    \iden &= \prod_{i=1}^{\abs{\mathcal P}} \gamma_{i}^2 \\
    % &= -i^{\abs{\mathcal P}} \prod_{i=1}^{\abs{\mathcal P}} -i\gamma_{i}\gamma_{i+1}\\
    &=-i^{\abs{\mathcal P}} \prod_{i=1}^{\abs{\mathcal P}}  A_{i,i+1} \equiv G 
\end{split}
\end{equation}
This relation will later yield a constraint on the qudit operators.
We will have to prepare the system in a subspace of the total Hilbert space, for which we have that $G\ket\psi = +\ket\psi$ or, shortly: $G \eqconstr \iden$. This is trivially fulfilled on the fermionic side, but yields non-trivial constraints on the qudit side.

\section{Local fermion-to-qudit mappings}

\subsection{Ququarts operators}
We focus on ququarts, namely qudits with four levels. This choice is motivated by the fact that four levels allow for encoding both the fermionic occupation number and parity information, or alternatively, the four occupation states $\{\ket{0}, \ket{\uparrow},\ket{\downarrow}, \ket{\uparrow, \downarrow}\}$. As we will demonstrate, this allows the construction of strictly local Hamiltonians without the overhead of non-local parity strings, which are typically required in qubit encodings, and a fermion-to-qudit ratio of $1$.

Here, the Euclidean Dirac matrices play a similar role as the Pauli matrices for qudits, as in Ref.~\cite{Vezvaee2024}.
These $\Gamma$ matrices introduce a representation of the Clifford algebra~\cite{Catto2013,Wang2020,Cui2015} and obey the anti-commutation relations $\{\Gamma^\mu,\Gamma^\nu\}=2\delta_{\mu\nu}$.
A fifth anti-commuting operator is obtained as:
\begin{equation}
    \tilde\Gamma = -\Gamma^1\Gamma^2\Gamma^3\Gamma^4
\end{equation}
which anti-commutes with the other operators.
For brevity, we will also introduce the notations
\begin{align}
\Gamma^{\mu \nu} &= \Gamma^\mu \Gamma^\nu \\
\tilde{\Gamma}^{\mu} &= \tilde{\Gamma} \Gamma^\mu
\end{align}
Note that, $\Gamma^\mu$ and $i\tilde{\Gamma}^\mu$ are hermitian operators. We provide explicit representations in \cref{app:gamma_matrices}.
%The latter have the anti-commutation relations 
%\begin{equation}
%    \{\Gamma^{\mu}, \Gamma^{\nu}\}=-2\delta_{\mu\nu}
%\end{equation}
%and the useful properties 
%\begin{align}
%    [\Gamma^\mu , \tilde{\Gamma}^{\nu}] & = -2\tilde \Gamma \delta_{\mu\nu} \\
%    \{\tilde\Gamma, \tilde{\Gamma}^{\mu}\} &= 0
%\end{align}

\subsection{Spinless fermions}\label{spinless}
%\subsubsection{Operator identification}
We will now consider a 2D square lattice of size $L_x \times L_y$ and identify the fermionic modes with a given lattice position $\vec{r}$. 
We consider open boundaries and refer to Appendix~\ref{sec:pbc} for details about periodic boundary conditions.
Our goal is to introduce a representation of the edge and vertex operators in terms of the $\Gamma$ matrices above. We use the notation $\Gamma^\mu_{\vec{r}}$ to indicate the $\Gamma^\mu$ operator applied to the fermionic mode corresponding to site $\vec{r}$, while applying identity operators to the other modes.
These operators therefore obey the additional commutation relations $[\G{\mu}{\vec r}, \G{\nu}{\vec r'}] = 0$ when $\vec r \neq \vec r'$. 

\begin{figure}[H]
    \centering
    \includegraphics[width=0.7\linewidth]{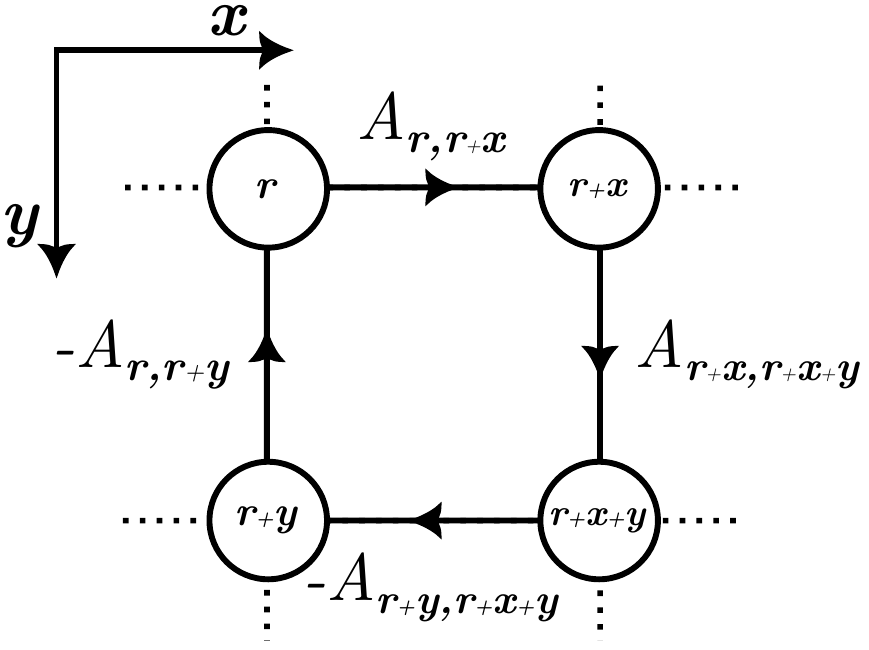}
\caption{Elementary plaquette of four sites, where the direction of the $A$ operator is highlighted. In particular, operators $A_{\vec{r+y,r}}$ and $A_{\vec{r+x+y,r+y}}$ are obtained using Eq.~\eqref{eq:antisymmetry_direction}.}
    \label{fig:elementary_plaquette}
\end{figure}

We can generate the required commutation relations through the identifications
\begin{align}
    A_{\vec r,\vec r + \vec x} = \G{1}{\vec r}\G{2}{\vec{r+x}}\\
    A_{\vec r,\vec r + \vec y} = \G{3}{\vec r}\G{4}{\vec{r+y}}
\end{align}
and impose that the operators in the negative $-\vec{x}$ and $-\vec{y}$ directions follow from the above definitions in the positive directions, using the anti-symmetry requirements on $A$, e.g.
\begin{equation}
A_{\vec{r}+\vec{y}, \vec{r}} := - A_{\vec{r}, \vec{r}+\vec{y}} = - \G{3}{\vec r}\G{4}{\vec{r+y}}\label{eq:antisymmetry_direction}
\end{equation}
The direction of the $A$ operators is shown in \cref{fig:elementary_plaquette} for an elementary plaquette, i.e.\ the smallest loop on a lattice.

We also identify the parity operator as
\begin{equation}
    B_{\vec r} = \tG{\vec{r}}
\end{equation}
It is straightforward to verify that with these definitions all (anti-)commutation relations of $A$ and $B$ in Section~\ref{sec:fermionic_operators} hold.

While this choice yields a simple and fully local realization of the fermionic algebra, other operator assignments are possible. As in the qubit case alternative identifications of edge and vertex operators may lead to mappings with different operator weights or gate requirements, while preserving locality and algebraic consistency.
Here we adopt a straightforward constructive approach: we choose the qudit-representation of the $A$ and $B$ operators based on a qudit lattice that resembles the fermionic one, and restricting to a fermion-to-qudit ratio of $1$. By identifying the parity operator $B$ with the fifth $\Gamma$ matrix, and assigning a $\Gamma$ operator per edge-attachment point per node, we satisfy the algebraic constraints and ensure the correct (anti-)commutation relations.
For the spinful case, we demonstrate how different qudit representations lead to different qudit-operator weights.

The hopping operators now follow
\begin{align}
    S_{\vec r,\vec {r+x}} &= -i \G{1}{\vec r}\G{2}{\vec{r+x}}\tG{\vec{r+x}} \label{eq:spinless_x} \\
    S_{\vec r,\vec {r+y}} &= -i \G{3}{\vec r}\G{4}{\vec{r+y}}\tG{\vec{r+y}} \label{eq:spinless_y}
\end{align}
A pair of fermions can be created on a neighboring set of lattice vertices using
\begin{equation} \label{eq:spinless_create}
f_{\vec{r}}^\dagger f_{\vec{r}+\vec{x}}^\dagger  
= \frac{i}{4} \G{1}{\vec{r}} \G{2}{\vec{r}+\vec{x}} (\iden + \tG{\vec{r}})(\iden + \tG{\vec{r}+\vec{x}})
\end{equation}
Notice that the two last factors guarantee that occupied sites cannot be occupied twice (and hence should be irrelevant when applied to the vacuum state). Along the $y$-direction we get a similar operator using $\Gamma^1\rightarrow\Gamma^3$ and $\Gamma^2\rightarrow\Gamma^4$.
Lastly, we focus on constraints that follow from Eq.~\eqref{eq:identity_prod_A}, that need to be satisfied in this mapping.
We therefore apply our fermion-to-qudit mapping defined above to the edge operators $A$ in $G$ in Eq.~\eqref{eq:identity_prod_A} and obtain
\begin{equation}\label{eq:spinless_const_plaquette}
\iden \eqconstr \G{13}{\vec{r}}\G{23}{\vec{r+x}}\G{42}{\vec{r+x+y}}\G{14}{\vec{r+y}} 
\end{equation}
In the above, we considered an elementary plaquette, i.e.\ $\abs{\mathcal{P}} = 4$. This constraint must be fulfilled exactly by any valid state and reduces the Hilbert space.

In addition, note that this constraint is essentially identical to the ground-state condition of the $\mathbb{Z}_2$ Toric code, which has been thoroughly studied for its usefulness in applications and for its state preparation, which can be performed in a reasonably efficient manner. See \cref{sec:toric_code} for further details.

\subsection{Spinful fermions}

%\subsubsection{General considerations}
We now identify each fermionic node to a combination of a position vector $\vec{r}$ and spin projection $\sigma \in \{\uparrow, \downarrow\}$. We only define  the $A$ edge operators between fermionic modes with the same spin projections. Notice that the general commutation relations require that for different spins $\sigma \neq \overline{\sigma}$ (with an abuse of notation)
\begin{align}
    \left[A_{\vec i, \vec j;\sigma}, A_{\vec k, \vec l; \overline{\sigma}}\right] = 0 \label{eq:comm_rel_1}\\
    \left[B_{\vec i;\sigma}, B_{\vec j;\overline{\sigma}}\right] = 0 \label{eq:comm_rel_2}
\end{align}
for all lattice vertices $\vec i, \vec j, \vec k, \vec l$. We will introduce two possible fermion-to-qudit identifications, yielding very different qudit Hamiltonians and constraints, and therefore different gate requirements. Therefore, we refrain from considering fermionic operators of both spin systems together, allowing us to bosonize also per spin sector individually.

% \subsection{Case 1: one qudit per spin type (``Spin-split mapping'')}
\subsubsection{Case 1: ``Spin-split mapping''}
\label{spinful_1}
Our first mapping uses the fermion-to-qudit operators introduced in the spinless case and generalizes them to the spinful case. Indeed, if we associate a separate qudit system to each of the spin sectors, the commutation relations in Eqs.~\eqref{eq:comm_rel_1}-\eqref{eq:comm_rel_2} are trivially fulfilled, since the $\Gamma$ operators on two qudits on different lattices commute. Therefore, we will introduce a secondary lattice composed of ququarts that we will indicate using the~$'$ indicator on the spatial index, i.e.\ $\vec{r}'$. We will refer to this mapping as the spin-split mapping, referring to the parallel treatment of the spin sectors.
For the parity factors, we choose
\begin{align}
    B_{\vec{r}; \uparrow} &= \tilde{\Gamma}_{\vec{r}} \\
    B_{\vec{r}; \downarrow} &= \tilde{\Gamma}_{\vec{r}'} 
\end{align}
For the link operators we take
\begin{align}
    A_{\vec{r},\vec{r}+\vec{x};\uparrow} &=  \Gamma^{1}_{\vec{r}} \Gamma^{2}_{\vec{r}+\vec{x}} \\
    A_{\vec{r},\vec{r}+\vec{x};\downarrow} &=  \Gamma^{1}_{\vec{r}'} \Gamma^{2}_{\vec{r}'+\vec{x}}
\end{align}
and the same for the $\vec{y}$ direction
\begin{align}
    A_{\vec{r},\vec{r}+\vec{y};\uparrow} &=  \Gamma^{3}_{\vec{r}} \Gamma^{4}_{\vec{r}+\vec{y}} \\
    A_{\vec{r},\vec{r}+\vec{y};\downarrow} &=  \Gamma^{3}_{\vec{r}'} \Gamma^{4}_{\vec{r}'+\vec{y}}
\end{align}
The link operators in the opposite direction are again defined from the above definitions by exploiting the anti-symmetry of $A$. We now obtain the hopping terms
\begin{align}
    S_{\vec{r}, \vec{r}+\vec{x}; \uparrow} + S_{\vec{r}+\vec{x}, \vec{r}; \uparrow} &= i \Gamma^{1}_{\vec{r}} \Gamma^{2}_{\vec{r}+\vec{x}} (\tilde{\Gamma}_{\vec{r}} - \tilde{\Gamma}_{\vec{r}+\vec{x}}) \\
    S_{\vec{r}, \vec{r}+\vec{x}; \downarrow} + S_{\vec{r}+\vec{x}, \vec{r}; \downarrow} &= 
    i \Gamma^{1}_{\vec{r}'} \Gamma^{2}_{\vec{r}'+\vec{x}} (\tilde{\Gamma}_{\vec{r}'} - \tilde{\Gamma}_{\vec{r}'+\vec{x}})
\end{align}
The plaquette constraints for both spins are independent as well. For example, for $\sigma = \uparrow$, we obtain
\begin{equation}
\begin{split}
    &A_{\vec{r}, \vec{r}+\vec{x}; \uparrow} A_{\vec{r}+\vec{x}, \vec{r}+\vec{x}+\vec{y}; \uparrow} A_{\vec{r}+\vec{x}+\vec{y}, \vec{r}+\vec{y}; \uparrow} A_{\vec{r}+\vec{y}, \vec{r}; \uparrow} \\
    % &= A_{\vec{r}, \vec{r}+\vec{x}; \uparrow} A_{\vec{r}+\vec{x}, \vec{r}+\vec{x}+\vec{y}; \uparrow} A_{\vec{r}+\vec{y},\vec{r}+\vec{x}+\vec{y}; \uparrow} A_{\vec{r}, \vec{r}+\vec{y}; \uparrow} \\
    &= - \G{13}{\vec{r}} \G{23}{\vec{r}+\vec{x}} \G{24}{\vec{r}+\vec{x}+\vec{y}} \G{14}{\vec{r}+\vec{y}} 
\end{split}
\end{equation}
such that the constraint reads
\begin{align}
 \iden &\eqconstr \Gamma^{13}_{\vec{r}} \Gamma^{23}_{\vec{r}+\vec{x}} \Gamma^{42}_{\vec{r}+\vec{x}+\vec{y}} \Gamma^{14}_{\vec{r}+\vec{y}} 
\end{align}
(and similarly for the second qudit system), which is the same as for the spinless case in \cref{eq:spinless_const_plaquette}. 
%Hence, this can similarly be re-written as
%\begin{align}
%    \iden &\eqconstr \G{23}{\vec{r}}\G{23}{\vec{r+x}}\G{23}{\vec{r+x+y}}\G{23}{\vec{r+y}} \tilde\Gamma_{\vec {r}}\tilde\Gamma_{\vec {r+x+y}}
%\end{align}

\begin{figure*}
    \centering
    \includegraphics[width=0.9\linewidth]{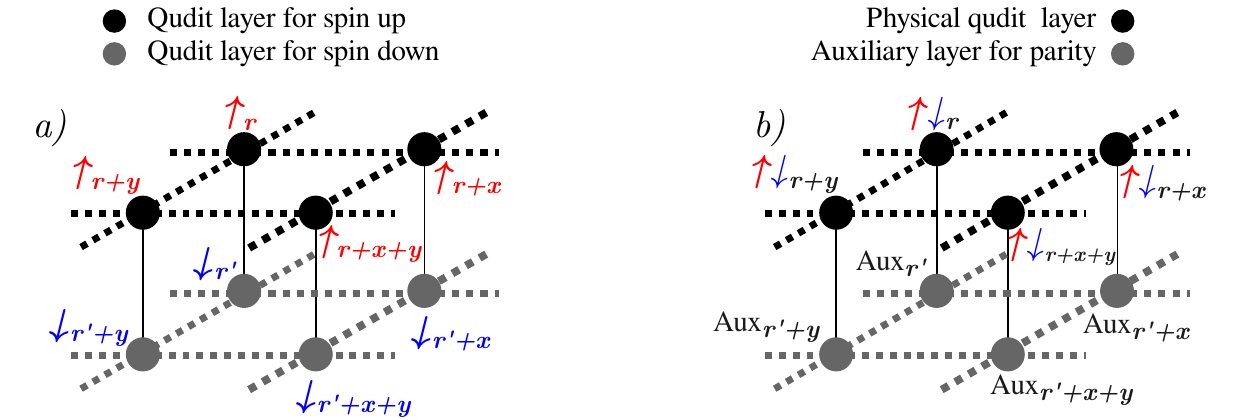}
    \caption{Depiction of the two spinful fermion-to-qudit mappings. In the spin-split mapping (left, panel a), the upper layer decodes all the information for spin-$\uparrow$ fermions while the bottom layer encodes all the information on spin-$\downarrow$ fermions. In the auxiliary-parity mapping (right, panel b), the upper layer contains partial information on the fermions of both spin species, while the bottom layer serves as an auxiliary system to store the parity and localize the qudit operators.}
    \label{fig:two_mappings}
\end{figure*}

% \subsection{Case 2: a qudit for the auxiliary modes (``Auxiliary-parity mapping'')}\label{spinful_2}
\subsubsection{Case 2: ``Auxiliary-parity mapping''}\label{spinful_2}
In the first qudit mapping, we used a qudit to encode spin-$\uparrow$ and a qudit for spin-$\downarrow$. 
In this section, we use a qudit to encode the physical information of both spin-$\uparrow$ and spin-$\downarrow$, and introduce an auxiliary set of qudits to maintain the corresponding commutation relations, i.e.\ to store the parity. We will refer to this mapping as the auxiliary-parity mapping, referring to the use of qudits as dominantly parity storage.
This mapping will be equivalent to localizing the qudit-generalized Jordan-Wigner strings by introducing an auxiliary qudit for each physical qudit and redefining the order of the sites by placing one auxiliary after every physical one as in the qubit case proposed in Ref.~\cite{verstraete2005mapping}. 
A recent and closely related work~\cite{Chizzini2024} independently applies this localization strategy in the qudit setting, extending the auxiliary fermion method \cite{verstraete2005mapping,ball2005fermions} to higher-dimensional local Hilbert spaces. Their construction, though distinct in formalism, achieves the same fermion-to-ququart mapping. Both spinful mappings introduced in our work are illustrated in \cref{fig:two_mappings}.

We start by association $\Gamma^1$ and $\Gamma^2$ to spin-$\uparrow$, while $\Gamma^3$ and $\Gamma^4$ will correspond to spin-$\downarrow$.
We then define the $B$ parity operators
\begin{align}
    B_{\vec r ; \uparrow} &= i \G{1}{\vec r}\G{2}{\vec r} \\
    B_{\vec r ; \downarrow} &= i \G{3}{\vec r}\G{4}{\vec r}
\end{align}
while edge operators become
\begin{equation}
\begin{split}
        A_{\vec r , \vec{r+x}; \uparrow} &=
        -i \tilde \Gamma_{\vec r} \tilde\Gamma_{\vec r'} \G{2}{\vec r}\G{2}{\vec{r+x}}\\
        &= -i\tilde\Gamma^{2}_{\vec r}\tilde\Gamma_{\vec r'} \G{2}{\vec{r+x}}
\end{split}
\end{equation}
\begin{equation}
\begin{split}
        A_{\vec r , \vec{r+y}; \uparrow} &=
        \tilde \Gamma_{\vec r} \tilde\Gamma_{\vec {r+y}} \G{2}{\vec r}\G{2}{\vec{r+y}}\G{1}{\vec r'} \G{2}{\vec {r'+y}}\\
        &= \tilde\Gamma^{2}_{\vec r}\tilde\Gamma^{2}_{\vec {r+y}} \G{1}{\vec{r'}} \G{2}{\vec{r'+y}}
\end{split}
\end{equation}
Spin-$\downarrow$ operators can be derived exchanging $\Gamma^1\rightarrow \Gamma^3$ and $\Gamma^2\rightarrow \Gamma^4$. All the $A$ and $B$ operators are correctly hermitian, while the operators in the opposite direction are again defined by anti-symmetry. Hence, the above representation is guided by the choice of the parity operator $B$, where we store the parity of both spins in the first and last two levels, respectively. The link operators $A$ then naturally follow from the requirement to fulfil the (anti)-commutation relations.

We obtain the horizontal hopping terms
\begin{align}
    S_{\vec{r}, \vec{r}+\vec{x}; \uparrow} + S_{\vec{r}+\vec{x}, \vec{r}; \uparrow} &=
    i \tilde \Gamma_{\vec r}\tilde \Gamma_{\vec r'}(\G{1}{\vec r}\G{2}{\vec{r+x}} - \G{2}{\vec r}\G{1}{\vec{r+x}})
\end{align}
while the vertical hopping operators, analogously become
\begin{align}
    &S_{\vec{r}, \vec{r}+\vec{y}; \uparrow} + S_{\vec{r}+\vec{y}, \vec{r}; \uparrow}  \nonumber \\
    &= \tilde \Gamma_{\vec r}\tilde \Gamma_{\vec{r+y}} (\G{1}{\vec r}\G{2}{\vec{r+y}} - \G{2}{\vec{r}}\G{1}{\vec {r+y}})\G{1}{\vec r'}\G{2}{\vec{r'+y}}
\end{align}

Finally, we can derive the elementary plaquette constraints.
In particular, for $\sigma=\uparrow$:
\begin{align}
    \iden &\eqconstr (
    \tilde\Gamma_{\vec{r+x}}
    \tilde\Gamma_{\vec{r+x+y}}
    )(
    \tilde\Gamma^{1}_{\vec r'}
    \G{1}{\vec{r'+x}}
    \G{2}{\vec{r'+x+y}}
    \tilde\Gamma^{2}_{\vec {r'+y}}
    )
\end{align}

Notice that this mapping yields operators that appear very similar to the forms obtained in the BVC mapping for qubits, or the mapping in Ref.~\cite{po2021symmetric}. 

% \section{State preparation}
% Consider the spinless case where we need to prepare a state that fulfills the plaquette constraints
% \begin{equation}
% \G{13}{\vec{r}}\G{23}{\vec{r+x}}\G{42}{\vec{r+x+y}}\G{14}{\vec{r+y}} = \iden
% \end{equation}
% We either follow the recipe from my paper, or we take a different route. We connect the local mapping to the Jordan-Wigner transformed case, where the vacuum is trivial. 
% % https://www.jmilne.org/not/Mxymatrix.pdf

%\newpage
\section{Applications on 2D lattices}
To demonstrate and validate our mappings in practice, we simulate prototypical fermionic lattice Hamiltonians in 2D using the abovementioned mappings. First, we transform the fermionic operators into qudit operators (i.e.\ $\Gamma$ matrices) for a given mapping.
Next, we prepare the vacuum state, ensuring it satisfies all local constraints $G \overset{c}{=} \iden$. Therefore, we apply projector operators to an initial state, $P_p = \frac{\iden - G_p}{2}$ for each plaquette $p$, i.e.\ $P = \bigotimes_p P_p$. We will demonstrate in Section~\ref{sec:toric_code} that this is equivalent to preparing the ground state of the $\mathbb{Z}_2$ Toric ground state.
Once the vacuum is prepared, we use the $f^\dagger_i f^\dagger_j$ operators to insert localized fermion pairs.
Finally, we simulate the quench dynamics of this state by performing a Trotterization of the qudit Hamiltonian. The code used for this section is available on GitHub~\cite{github}.

\subsection{$t$-$V$ model dynamics}
We start by simulating the spinless $t$-$V$ model using the mapping introduced in \cref{spinless}. The spinless $t$-$V$ Hamiltonian transforms as
\begin{equation}
\begin{split}
    H =& -T \sum_{\langle \vec{r, s}\rangle} (f_{\vec r}^\dagger f_{\vec s} + f_{\vec s}^\dagger f_{\vec r}) + V \sum_{\langle \vec{r,s}\rangle} n_{\vec r} n_{\vec s} \\ 
    \to& -iT \sum_{\langle \vec{r,r+x}\rangle} \Gamma^1_{\vec r}\Gamma^2_{\vec{r+x}}(\tilde\Gamma_{\vec r} - \tilde\Gamma_{\vec{r+x}}) \\
    &- iT \sum_{\langle \vec{r,r+y}\rangle} \Gamma^3_{\vec r}\Gamma^4_{\vec{r+y}}(\tilde\Gamma_{\vec r} - \tilde\Gamma_{\vec{r+y}}) \\ 
    &+ \frac{V}{4} \sum_{\langle \vec{r,s}\rangle} (I- \tilde\Gamma_{\vec r})(I-\tilde\Gamma_{\vec s}) \label{eq:H_qudit_tV} \\
\end{split}
\end{equation}
with $\langle \vec r,\vec s\rangle$ representing all nearest neighbor edges, and $\langle \vec{r,r+x}\rangle$ and $\langle \vec{r,r+y}\rangle$ the horizontal and vertical edges respectively.

We consider a $2 \times 3$ lattice ($N=6$), where each site can either be unoccupied ($0$) or occupied by a particle ($1$). The fermionic state we aim to prepare reads
\begin{equation*}
\ket{\psi_0}=\frac{1}{\sqrt{2}}\left(\ket{\begin{matrix}
1 & 1 & 0\\
0 & 0 & 0
\end{matrix}} + \ket{\begin{matrix}
1 & 0 & 1\\
0 & 0 & 0
\end{matrix}}\right)
\end{equation*}
where the rows and columns relate to the position on the lattice. This initial state exhibits both non-zero kinetic and potential energy, and we explore interference and correlations generated during its dynamics.
To this end, on the qudit circuit, we start from an initial $\ket{0}^{\otimes N}$ for the $N=6$ qudits. From this, we construct the non-trivial vacuum state that fulfills the constraint in Eq.~\eqref{eq:spinless_const_plaquette} by applying the corresponding mapped projection operator to the initial state.  After, we implement the desired state by applying an operator built with a combination of the pair-creation and hopping operators introduced in \cref{eq:spinless_x,eq:spinless_y,eq:spinless_create}. 
In this case this will correspond to applying the fermionic operator $O = \frac{1}{\sqrt{2}}\left(S_{\vec{r+x,r+2x}}f^\dagger_{\vec r}f^\dagger_{\vec{r+x}} + f^\dagger_{\vec r} f^\dagger_{\vec{r+x}}\right)$ to the vacuum state.

To implement the Trotterized dynamics, we group the terms in the Hamiltonian into groups of non-commuting terms. We therefore separate even and odd lattice edges, forming two groups (A and B). All operators within a group commute, allowing us to apply them in parallel.
The ordering of terms is structured as follows: first horizontal hopping terms ($H^{hop,x}$) of even and odd edges, followed by vertical hopping terms ($H^{hop,y}$) on even and odd edges, and then all the interaction terms ($H^{int}$). This yields the following set of non-commuting operators~\cite{campbell2014enhanced}.
\begin{equation}
\begin{split}
H_{set} = \{H^{hop,x}_A, H^{hop,x}_B, H^{hop,y}_A, 
H^{hop,y}_B, H^{int}\}\label{eq:Hset}
\end{split}
\end{equation}
We symmetrize to obtain a second-order Trotterization scheme, i.e.\
\begin{equation}
\begin{split}
e^{-iHt} \approx \left(\prod_{j=1}^{m} e^{-iH_j \tau/2} \prod_{j=m}^{1} e^{-iH_j \tau/2}\right)^n
\end{split}
\end{equation}
where $m$ is the number of terms in the ordered set $H_{set}$ (i.e.\ $m=5$ in \cref{eq:Hset}) and $\tau=t/n$ is the Trotterization time step.

\begin{figure}[t]
    \centering
    \includegraphics[width=\linewidth]{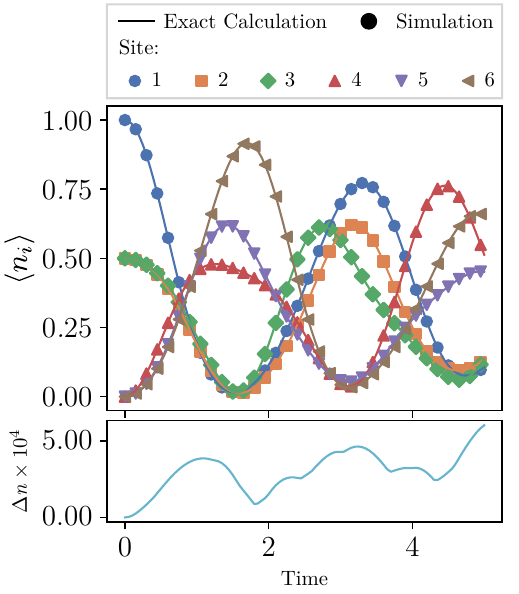}
    \caption{Evolution of the site occupation of an initial state under the $t$-$V$ Hamiltonian on a $2 \times 3$ lattice. Sites are numbered left-to-right, top-to-bottom. Solid lines show the results obtained with exact diagonalization, while markers represent the qudit-based Trotter simulation. The total error is the absolute sum of the occupation number deviations at each site between the simulated and the classically computed values. The parameters used for the simulation are $\tau=0.05/T$ and $V/T=0.5$.}
    \label{fig:spinless_evolution}
\end{figure}

In \cref{fig:spinless_evolution} we show for all sites in the lattice the real-time evolution of the site occupation. In this mapping, the latter is represented by: $\expval{n_{\vec r}}=\expval{\frac{I-\tilde\Gamma_{\vec r}}{2}}$.
We show the total error on the occupations $\Delta n = \sum_{\vec r} \abs{\expval{n_{\vec r}} - \tilde n_{\vec r}}$ for all sites. Here, $\tilde n_{\vec r}$ is the exact result obtained through exact diagonalization of the fermionic Hamiltonian (using NetKet~\cite{Vicentini2022}). This demonstrates that the mapping reproduces the correct dynamics.

\subsection{Fermi-Hubbard quantum dynamics}
\begin{figure}[tp]
    \centering
    \includegraphics[width=\linewidth]{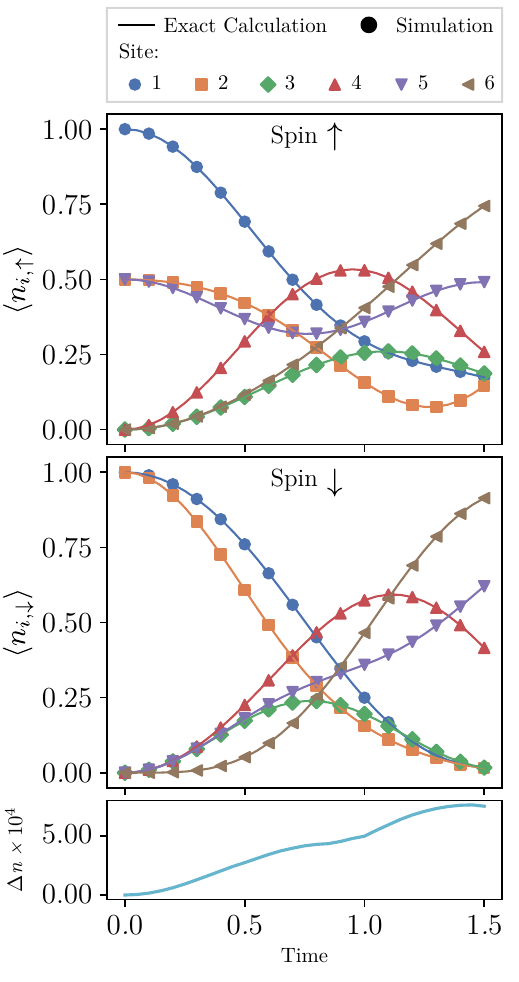}
    \caption{Evolution of site densities of an initial state subject to the dynamics governed by the Fermi-Hubbard model on a $2\times 3$ lattice. Sites are numbered left-to-right, top-to-bottom. Solid lines show the results obtained with exact diagonalization, while markers represent the qudit-based simulation based on the first spinful fermion-to-qudit mapping. The error on the bottom panel is the sum of the errors per spin sector. The parameters used in the dynamics are $\tau=0.05/J$ and $U/J=0.5$.
    }
    \label{fig:spinful_evolution}
\end{figure}

We apply a similar procedure to validate the dynamics using the Fermi-Hubbard Hamiltonian
\begin{equation}
    H = -J\sum_{\substack{\langle \vec{r, s} \rangle \\ \sigma\in\{\uparrow,\downarrow\}}} (f^\dagger_{\vec{r};\sigma} f_{\vec{s};\sigma} + f^\dagger_{\vec{s};\sigma} f_{\vec{r};\sigma}) + U\sum_{\vec{r}} n_{\vec{r};\uparrow}n_{\vec{s};\downarrow} .
\end{equation}
We again consider a $2\times 3$ lattice, but now incorporate the two spin species.
Using the first spinful mapping presented in \cref{spinful_1} we obtain:
\begin{equation}
\begin{split}
    H \to & -iJ \sum_{\langle \vec{r,r+x}\rangle} \Gamma^1_{\vec r}\Gamma^2_{\vec{r+x}}(\tilde\Gamma_{\vec{r}} - \tilde\Gamma_{\vec{r+x}}) \\
    &- iJ \sum_{\langle \vec{r,r+x}\rangle} \Gamma^1_{\vec{r'}}\Gamma^2_{\vec{r'+x}}(\tilde\Gamma_{\vec{r'}} - \tilde\Gamma_{\vec{r'+x}}) \\
    &- iJ \sum_{\langle \vec{r,r+y}\rangle} \Gamma^3_{\vec r}\Gamma^4_{\vec{r+y}}(\tilde\Gamma_{\vec r} - \tilde\Gamma_{\vec {r+y}}) \\
    &- iJ \sum_{\langle \vec{r,r+y}\rangle} \Gamma^3_{\vec{r'}}\Gamma^4_{\vec{r'+y}}(\tilde\Gamma_{\vec{r'}} - \tilde\Gamma_{\vec{r'+y}}) \\
    &+ \frac{U}{4} \sum_{\vec{r}} (I-\tilde \Gamma_{\vec r})(I-\tilde\Gamma_{\vec{r'}}) \label{eq:H_qudit_FH}
\end{split}
\end{equation}
This mapping is the generalization of the spinless case. However, the Hamiltonian differs in that the potential energy now arises from interactions between two spin modes at the same site, rather than from different sites. On the fermionic side the local Hilbert space per site reads $\{\ket{0}, \ket{\uparrow}, \ket{\downarrow}, \ket{\uparrow\downarrow}\}$, corresponding respectively to an empty site, occupation by a single up-fermion ($\uparrow$), a single down-fermion ($\downarrow$), or doubly occupied ($\uparrow\downarrow$). We will evolve the initial state that on the fermionic side corresponds to 
\begin{equation*}
    \ket{\psi_0} = \frac{1}{\sqrt{2}}\left(\ket{\begin{matrix}\uparrow\downarrow & \downarrow & 0 \\ 
    0 & \uparrow & 0 \end{matrix}} + \ket{\begin{matrix}\uparrow\downarrow & \uparrow\downarrow & 0 \\ 
    0 & 0 & 0 \end{matrix}}\right)
\end{equation*}
A similar Trotterization procedure as in the spinless case is used.
The numerical results are shown in \cref{fig:spinful_evolution}. 

We verify also the auxiliary-parity spinful mapping, which transforms the Hamiltonian into
\begin{equation}
\begin{split}
    &H \to \\
    & -iJ \sum_{\langle \vec{r,r+x}\rangle} \tilde\Gamma_{\vec r} \tilde\Gamma_{\vec{r'}} (\Gamma^1_{\vec r}\Gamma^2_{\vec{r+x}} - \Gamma^2_{\vec{r}}\Gamma^1_{\vec{r+x}}) \\
    &- iJ \sum_{\langle \vec{r,r+x}\rangle} \tilde\Gamma_{\vec r} \tilde\Gamma_{\vec{r'}} (\Gamma^3_{\vec r}\Gamma^4_{\vec{r+x}} - \Gamma^4_{\vec{r}}\Gamma^3_{\vec{r+x}}) \\
    &- J  \sum_{\langle \vec{r,r+y}\rangle} \tilde\Gamma_{\vec r} \tilde\Gamma_{\vec{r+y}}(\Gamma^1_{\vec r}\Gamma^2_{\vec{r+y}} - \Gamma^2_{\vec{r}}\Gamma^1_{\vec{r+y}})\Gamma^1_{\vec{r'}}\Gamma^2_{\vec{r'+y}} \\
    &- J  \sum_{\langle \vec{r,r+y}\rangle} \tilde\Gamma_{\vec r} \tilde\Gamma_{\vec{r+y}}(\Gamma^3_{\vec r}\Gamma^4_{\vec{r+y}} - \Gamma^4_{\vec{r}}\Gamma^3_{\vec{r+y}})\Gamma^3_{\vec{r'}}\Gamma^4_{\vec{r'+y}} \\
    &+ \frac{U}{4} \sum_{\vec r} (I-i\Gamma^1_{\vec r}\Gamma^2_{\vec r})(I-i\Gamma^3_{\vec r}\Gamma^4_{\vec r}) \label{eq:H_qudit_FH_2}
\end{split}
\end{equation}
Although this alternative mapping yields results comparable to the previous one, with similar errors in $\Delta n$, it introduces Hamiltonian terms with an increased qudit weight, as will be further discussed in the next section.

\subsection{Quantum dynamics: gate counting}\label{sec:gate_comparison}
The hardware implementation of qudits is still in its early stages and, as a result, there is no standardized gate set available.
We therefore refrain from performing a detailed gate-count comparison to implement the time evolution across different mappings.
Instead, we follow a more general approach and estimate the number of two-qudit unitaries required to implement a first-order Trotter step, namely how to implement the evolution operators $e^{-i\theta H}$ where $\theta$ is the chosen time step, similarly to what is carried out in Ref.~\cite{cade2020strategies} for qubit mappings. Hereby, we ignore factors coming from non-communicating neighboring edges and consider the cost and depth to implement a vertical and a horizontal hopping operator and the interaction term, similar to Ref.~\cite{cade2020strategies}.
Specifically, we assume that any 2-qudit gate can be implemented with a gate depth $=1$, while further neglecting single-qudit rotations.
This assumption is driven by the fact that single-qudit operations are performed with accuracies much higher than two-qudit operations. Therefore, the latter are the main factor determining the total fidelity of a circuit~\cite{Fowler2012,daSilva2011}. This procedure, however, introduces a useful starting point for gate compilations to specific gate sets.

\paragraph{Spinless case}
For the spinless case, the horizontal hopping operators read
\begin{align*}
    S_{\vec{r,r+x}}+S_{\vec{r+x,r}}&=i\Gamma^1_{\vec r}\Gamma^2_{\vec{r+x}} (\tilde\Gamma_{\vec r} - \tilde\Gamma_{\vec{r+x}}) \\
    &= -i\left(\tilde\Gamma^1_{\vec r}\Gamma^2_{\vec{r+x}} - \Gamma^1_{\vec r}\tilde\Gamma^2_{\vec{r+x}} \right)
\end{align*}
Since the two terms commute, to implement the evolution of the hopping term over a time $\theta$, we can apply two subsequent two-qudit operators: $\exp\left\{-\theta{\tilde\Gamma^1_{\vec r}\Gamma^2_{\vec{r+x}}}\right\}$ and $\exp\left\{\theta{\Gamma^1_{\vec r} \tilde\Gamma^2_{\vec{r+x}}}\right\}$. 
The same transformation applies to the vertical hopping operators using $\Gamma^1\rightarrow\Gamma^3$ and $\Gamma^2\rightarrow\Gamma^4$.
Note that all these terms operate on the same qudits, therefore the number of two-qudit gates will also correspond to the depth of the circuit.

The interaction term is given by:
\begin{equation}
    n_{\vec r}n_{\vec s}=\iden - \tilde\Gamma_{\vec r} - \tilde\Gamma_{\vec s} + \tilde\Gamma_{\vec{r}}\tilde\Gamma_{\vec s} \quad (\vec{r} \neq \vec{s})
\end{equation}
Thus it involves a single two-qudit operator. 
To simulate the evolution, we apply the two single-qudit operations: $\exp\left\{i\theta\tilde\Gamma_{\vec r}\right\}$ and $\exp\left\{i\theta\tilde\Gamma_{\vec s}\right\}$ and the two-qudit operator $\exp\left\{-i\theta\tilde\Gamma_{\vec r}\tilde\Gamma_{\vec {s}}\right\}$. Hence, we again obtain a two-qudit gate count of $1$. The total two-qudit gate count for an evolution step, considering horizontal and vertical hopping terms as well as interaction terms is:
\begin{equation*}
    2+2+1=5
\end{equation*}

\paragraph{Spinful spin-split mapping}
For the spin-split mapping, the hopping terms are equivalent to the spinless case, but are duplicated to account for both spin species.
Since the $\uparrow$ and $\downarrow$ operators act on different qudits, they can be executed simultaneously, resulting in a total depth $4$ for all hopping terms as before, while the number of two-qudit gates is doubled. The interaction potential of the Fermi-Hubbard model describes on-site interactions. While its form appears different than in $t$-$V$ model on the fermionic side, it obtains a similar form in the qudit-mapped form when comparing \cref{eq:H_qudit_FH} to \cref{eq:H_qudit_tV}, thus contributing a two-qudit gate count of $1$.
Therefore, the two-qudit gate count for this mapping is:
\begin{equation*}
    2(2+2)+1=9
\end{equation*}
As pointed out earlier, since the hopping terms of both spin species can be implemented in parallel, the circuit depth is reduced to $5$ as in the spinless case.

\paragraph{Spinful auxiliary-parity mapping}

In the spinful auxiliary-partiy mapping, the horizontal hopping operator is written as:
\begin{equation}\label{eq:spinful_h_to_conj}
    \begin{split}
    &S_{\vec{r}, \vec{r}+\vec{x}; \uparrow} + S_{\vec{r}+\vec{x}, \vec{r}; \uparrow} \\
    &=i\tilde\Gamma_{\vec r} (\Gamma^1_{\vec r}\Gamma^2_{\vec{r+x}} - \Gamma^2_{\vec r}\Gamma^1_{\vec{r+x}})\tilde\Gamma_{\vec{r'}} \\&=i (\tilde\Gamma^1_{\vec r}\Gamma^2_{\vec{r+x}} - \tilde\Gamma^2_{\vec r}\Gamma^1_{\vec{r+x}})\tilde\Gamma_{\vec{r'}}
    %&= U^\dagger_{(\vec{r,r+x})}(\tilde\Gamma_{\vec r} + \tilde\Gamma_{\vec{r+x}}) U_{(\vec{r,r+x})} \tilde\Gamma_{{\vec{r'}}}
    \end{split}
\end{equation}
This circuit involves three-qudit operators. We decompose its evolution into two-qudit operators using unitary conjugation. 
Using the identity
\begin{align}\label{eq:exp_property}
    e^{A \otimes B} = V^\dagger e^{D \otimes B} V
\end{align}
where $A = V^\dagger D V$ with $D$ diagonal and $A$ hermitian.
We can write the operators on $\vec r$ and $\vec{r+x}$ as:
\begin{equation}
\begin{split}
    &\tilde\Gamma^1_{\vec r}\Gamma^2_{\vec{r+x}} - \tilde\Gamma^2_{\vec r}\Gamma^1_{\vec{r+x}} \\
    &=U^\dagger_{(\vec{r,r+x})} \, \left(\tilde\Gamma_{\vec r} + \tilde\Gamma_{\vec{r+x}}\right)\, U_{(\vec{r,r+x})}
\end{split}
\end{equation}
The explicit form of $U$ can be found in \cref{app:unitary_conj}.
After applying the two-qudit  operator $U$ to $\vec{r}$ and $\vec{r+x}$, we apply the evolution terms that involve the third qudit: $\exp\left\{ -i\theta \tilde\Gamma_{\vec r}\tilde\Gamma_{\vec{r'}} \right\}$ and $\exp\left\{ -i\theta \tilde\Gamma_{\vec {r+x}}\tilde\Gamma_{\vec{r'}} \right\}$. Since both commute, they can be applied sequentially. Finally, we uncompute the transformation $U$. 
The circuit is schematically represented in \cref{fig:circuit1}, where the two evolution operators are indicated as $R_1$ and $R_2$:

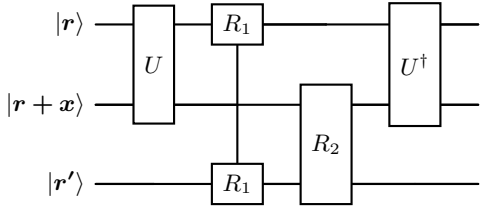
\begin{figure}[h!]
    \centering
    \begin{quantikz}
\lstick{\( \ket{\vec{r}}  \)} & \gate[2]{U} & \gate{R_1}               & \qw            & \gate[2]{U^\dagger} & \qw \\
\lstick{\( \ket{\vec{r+x}}\)} & \qw         & \qw                      & \gate[2]{R_2}  & \qw                 & \qw \\
\lstick{\( \ket{\vec{r'}} \)} & \qw         & \gate{R_1}\wire[u][2]{a} & \qw            & \qw                 & \qw
\end{quantikz}
    \caption{Schematic representation of the circuit implementing horizontal hopping in the second spinful mapping.}
    \label{fig:circuit1}
\end{figure}

As we can see, the horizontal hopping operators for the $\uparrow$ section can be performed with a circuit of $4$ two-qudit gates.
An additional equivalent circuit is required for the $\downarrow$ section. 

\begin{figure}[h!]
    \centering
\begin{quantikz}
\lstick{\( \ket{\vec r} \)}    & \gate[2]{V}  & \gate{R_1}               & \qw            & \gate[2]{V^\dagger} & \qw \\
\lstick{\( \ket{\vec{r+y}} \)}  & \qw         & \qw                      & \gate[2]{R_2}  & \qw                 & \qw \\
\lstick{\( \ket{\vec{r'}} \)}   & \gate[2]{W} & \gate{R_1}\wire[u][2]{a} & \qw            & \gate[2]{W^\dagger} & \qw \\
\lstick{\( \ket{\vec{r'+y}} \)} & \qw         & \qw                      & \qw            & \qw                 & \qw \\
\end{quantikz}
    
    \caption{Schematic representation of the two-qudit circuit implementing the vertical hopping in the second spinful mapping.}
    \label{fig:circuit2}
\end{figure}
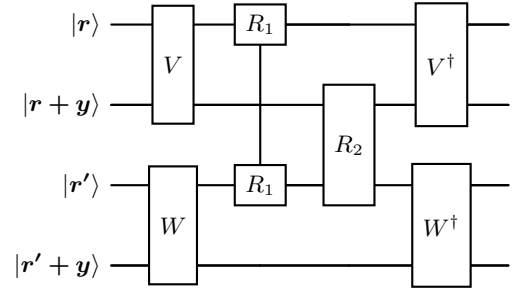

The vertical operators instead involve operation on four different qudits:
\begin{equation}\label{eq:spinful_v_to_conj}
\begin{split}
    &S_{\vec{r}, \vec{r+y}; \uparrow} + S_{\vec{r+y}, \vec{r}; \uparrow}\\
    &=\tilde\Gamma_{\vec r} \tilde\Gamma_{\vec{r+y}} (\Gamma^1_{\vec r}\Gamma^2_{\vec{r+y}} - \Gamma^2_{\vec r}\Gamma^1_{\vec{r+y}})\Gamma^1_{\vec{r'}}\Gamma^2_{\vec{r'+y}} \\
    &=(\tilde\Gamma^1_{\vec r}\tilde\Gamma^2_{\vec{r+y}} - \tilde\Gamma^2_{\vec r}\tilde\Gamma^1_{\vec{r+y}})\Gamma^1_{\vec{r'}}\Gamma^2_{\vec{r'+y}} 
\end{split}
\end{equation}
We apply again \cref{eq:exp_property} similarly to the horizontal hopping term to simplify the operators between brackets:
\begin{equation}
\begin{split}
&\tilde\Gamma^1_{\vec r}\tilde\Gamma^2_{\vec{r+y}} - \tilde\Gamma^2_{\vec r}\tilde\Gamma^1_{\vec{r+y}}\\
&=
V^\dagger_{(\vec{r,r+y})}\,\left(\tilde\Gamma_{\vec r} + \tilde\Gamma_{\vec{r+y}}\right)\,V_{(\vec{r,r+y})}
\end{split}
\end{equation}
After the transformation $V$, we still are left with weight three operators.
However, we can also optimize the operators on $\vec{r'}$ and $\vec{r'+y}$ with the same method, using \cref{eq:exp_property} and identifying the operators on $\vec{r}$ and $\vec{r+y}$ with $B$:
\begin{equation}
\Gamma^1_{\vec{r'}}\Gamma^2_{\vec{r'+y}}=W^\dagger_{(\vec{r',r'+y})}\,\tilde\Gamma_{\vec{r'}} \, W_{(\vec{r',r'+y})}
\end{equation}
In the above, $V$ and $W$ can be computed in parallel. We then apply the two resulting evolutions $\exp\left\{-i\theta \tilde\Gamma_{\vec{r}}\tilde\Gamma_{\vec{r'}}\right\}$ and $\exp\left\{-i\theta \tilde\Gamma_{\vec{r+y}}\tilde\Gamma_{\vec{r'}}\right\}$ and uncompute $V$ and $W$ again in parallel. As a result, we need $6$ two-qudit gates to implement the interaction term, and obtain a two-qudit depth of $4$ (see \cref{fig:circuit2}), similar to the BVC fermion-to-qubit mapping in Ref.~\cite{cade2020strategies}.

The interaction term in the second spinful mapping in \cref{eq:H_qudit_FH_2} involves a single qudit only, and therefore does not add to the two-qudit count. Overall, we obtain a two-qudit gate count of:
\begin{equation}
    2(4+6)=20
\end{equation}

\subsection{Operator-weight comparison}
\begin{table*}[bth]
    \centering
    \footnotesize
    \caption{Qudit-weight comparison for different mappings, including both spinless (top) and spinful (bottom) fermions. $d$ represents the qudit levels ($2$ for qubits, $4$ for ququarts). $W_x$ and $W_y$ refer to the qudit weights of horizontal and vertical hopping terms, respectively. $W_{int}$ denotes the qudit weight of interaction terms, and $W_g$ indicates the qudit-weight for the plaquette constraints. We also indicate the ratio of number of qudits per fermionic mode. We assume a square $L_x \times L_y$ lattice and the $t$-$V$ model (spinless) and Fermi-Hubbard model (spinful) interactions. The results for the mappings derived in this work are indicated in bold. For Jordan Wigner transformations (JWT), we assume a horizontal snake ordering.\label{tab:gate_count_table}}
    \begin{tabular}{l|c|c|c|c|c|c}
        \hline
        \textbf{Method} & $d$ & ratio &  $W_x$ &  $W_y$ & $W_{int}$ &  $W_g$ \\
        \hline
        JW   & $2$ & $1$ & $2$ & $\order{L_x}$ & $2$ & N/A  \\
        \textbf{Spinless local}  & $4$ & $1$ & $2$ & $2$ & $2$ & $4$  \\
        \hline
        Generalized JW (QFM)~\cite{Vezvaee2024} & $4$ & $\frac{1}{2}$ & $2$ & $\order{L_x}$ & $1$ & N/A \\
        BVC ~\cite{verstraete2005mapping}  & $2$ & $2$ & $3$ & $4$ & $2$ & $6$  \\        
        \textbf{Local spin split}  & $4$ & $1$ & $2$ & $2$ & $2$ & $4$  \\  
        \textbf{Spinful auxiliary parity}  & $4$ & $1$ & $3$ & $4$ & $1$ &  $6$ \\
        \hline
    \end{tabular}
    
\end{table*}
In the previous section, we introduced a more compact version of the qudit Hamiltonians, which is better for counting the operator weights. We compare our results with known fermion-to-qubit mappings in Table~\ref{tab:gate_count_table}. In the spinless case, we observe that local transformations have the same qudit-to-fermion-mode ratio as in the qubit Jordan-Wigner transformation, with the significant additional benefit of having all local hopping operators. In the spinful case, we observe that local transformations can implement the Fermi-Hubbard model with similar operator weights as in the spinless case. The second local spinful mapping yields operator weights similar to the BVC mapping on qubits, yet reduces the qudit-to-fermion-mode ratio and interaction term weights. Hence, the mapping can be seen as the BVC generalization to qudits. As observed also in the previous section, the first local spinful mapping yields a number of advantages, reducing the operator weight of the vertical hopping and the constraint operator compared to the BVC transformation on qubits.

\section{Connection to the $\mathbb{Z}_2$ Toric code}
\label{sec:toric_code}
\begin{figure*}[tbh]
    \centering
    \includegraphics[width=0.9\linewidth]{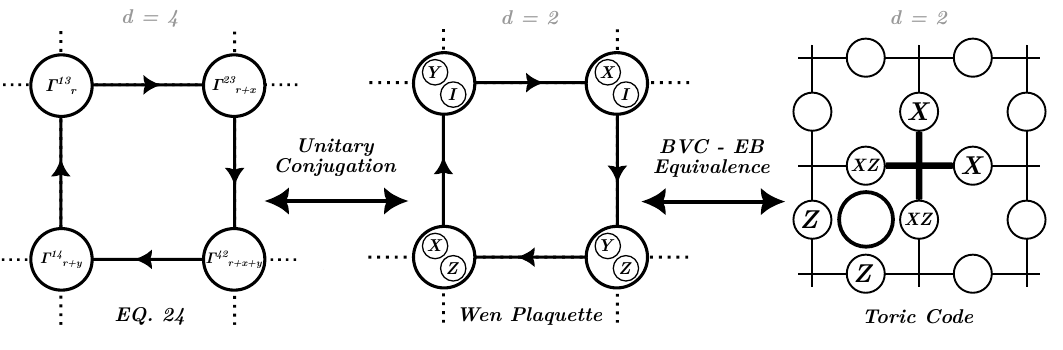}
    \caption{Depiction of the equivalence between the spinless plaquette operator of \cref{eq:spinless_const_plaquette} and the $\mathbb{Z}_2$ Toric code.
    The plaquette operator is transformed through local unitary conjugation in \cref{eq:spinless_plaquette_wenned} to the dynamical Wen plaquette model, which is equivalent to the Toric code, using the local unitary and node-to-edge transformation in Ref.~\cite{chen2023equivalence}.
    }
    \label{fig:wen_plaquette_equivalence}
\end{figure*}
Finally, we show that satisfying the plaquette constraint in Eq.~\eqref{eq:spinless_const_plaquette} is equivalent to preparing the ground state of the $\mathbb{Z}_2$ Toric ground state. This equivalence has also been demonstrated for all 2D fermion-to-qubit mappings in Ref.~\cite{chen2023equivalence}. The procedure is summarized in \cref{fig:wen_plaquette_equivalence}.
The constraint in \cref{eq:spinless_const_plaquette} can be written in terms of the Pauli matrices applied to two two-level (qubit) subspaces of the qudits. Assuming we decompose the ququarts into qubit pairs using the representation in App.~\ref{app:gamma_matrices}, we obtain
\begin{align}
    -i\Gamma^{13} &= - \sigma^y \otimes \sigma^x \\
    -i\Gamma^{23} &= \sigma^x \otimes \sigma^x \\
    -i\Gamma^{42} &= - \sigma^x \otimes \sigma^y \\
    -i\Gamma^{14} &= - \sigma^y \otimes \sigma^y
\end{align}
such that the plaquette constraint reads
\begin{multline}
    G_{\vec{r}} = -\left(\sigma^y \otimes \sigma^x\right)_{\vec{r}}\left(\sigma^x \otimes \sigma^x\right)_{\vec{r}+\vec{x}} \cdot \\ \cdot \left(\sigma^x \otimes \sigma^y\right)_{\vec{r}+\vec{x}+\vec{y}}\left(\sigma^y \otimes \sigma^y\right)_{\vec{r}+\vec{y}}
\end{multline}
By applying unitary conjugation on each individual term with $U=\textrm{CNOT}_{12}$, we obtain
\begin{align}
    U (\sigma^y \otimes \sigma^x) U^\dagger &= \sigma^y \otimes \iden_{2\times 2} \\
    U (\sigma^x \otimes \sigma^x) U^\dagger &= \sigma^x \otimes \iden_{2\times 2} \\
    U (\sigma^x \otimes \sigma^y) U^\dagger &= \sigma^y \otimes \sigma^z \\
    U (\sigma^y \otimes \sigma^y) U^\dagger &= -\sigma^x \otimes \sigma^z 
    % U (\sigma^y \otimes \sigma^x) U^\dagger &= \sigma^y \otimes \iden_{2\times 2} = i (\sigma^x \sigma^z)  \otimes \iden_{2\times 2} \\
    % U (\sigma^x \otimes \sigma^x) U^\dagger &= \sigma^x \otimes \iden_{2\times 2} \\
    % U (\sigma^x \otimes \sigma^y) U^\dagger &= \sigma^y \otimes \sigma^z = i(\sigma^x \sigma^z) \otimes \sigma^z \\
    % U (\sigma^y \otimes \sigma^y) U^\dagger &= -\sigma^x \otimes \sigma^z 
\end{align}
We introduce the shorthand notation
\begin{equation}
  \sigma_{\vec{r}}^{\mu, a} = \begin{cases}
    \sigma_{\vec{r}}^{\mu} \otimes \iden_{2\times 2}, & \text{if $a=1$}.\\
    \iden_{2\times 2} \otimes \sigma_{\vec{r}}^{\mu}, & \text{if $a=2$}.
  \end{cases}
\end{equation}
with $\mu=x, y, z$, such that the plaquette operator in Eq.~\eqref{eq:spinless_const_plaquette} reads
\begin{align}\label{eq:spinless_plaquette_wenned}
    G_v &= \sigma^{y,1}_{\vec{r}} \sigma^{x,1 }_{\vec{r}+\vec{x}} \sigma^{y,1}_{\vec{r}+\vec{x}+\vec{y}}  \sigma^{x,1}_{\vec{r}+\vec{y}} \sigma^{z,2}_{\vec{r}+\vec{x}+\vec{y}} \sigma^{z,2}_{\vec{r}+\vec{y}} %\\
    % &= \sigma^{x,1}_{\vec{r}} \sigma^{z,1}_{\vec{r}} \sigma^{x,1 }_{\vec{r}+\vec{x}} \sigma^{x,1}_{\vec{r}+\vec{x}+\vec{y}}  \sigma^{x,1}_{\vec{r}+\vec{y}} \sigma^{z,2}_{\vec{r}+\vec{x}+\vec{y}} \sigma^{z,2}_{\vec{r}+\vec{y}}
\end{align}
Hence, the plaquette constraints correspond to the dynamical Wen plaquette model 
\begin{align}
    \sigma^{y,1}_{\vec{r}} \sigma^{x,1 }_{\vec{r}+\vec{x}} \sigma^{y,1}_{\vec{r}+\vec{x}+\vec{y}}  \sigma^{x,1}_{\vec{r}+\vec{y}} &= \sigma^{z,2}_{\vec{r}+\vec{x}+\vec{y}} \sigma^{z,2}_{\vec{r}+\vec{y}}
\end{align}
with plaquette operators on the auxiliary system $a=1$ and a parity operator on the physical system $a=2$. See \cref{fig:wen_plaquette_equivalence} for reference.
This constraint is similarly obtained in the BVC fermion-to-qubit mapping, see Refs.~\cite{verstraete2005mapping, cade2020strategies, nys2023quantum}. The solution to this constraint can be obtained by exploiting the equivalence to the $\mathbb{Z}_2$ Toric code ground state, using the transformations provided in Refs.~\cite{chen2023equivalence, nys2023quantum} and explicitly shown in \cref{app:exact_bosonization}.
In particular, Ref.~\cite{chen2023equivalence} provides the mapping between the BVC mapping and the exact bosonization (EB) procedure~\cite{chen2018exact}, where the plaquette constraint of the latter corresponds to the $\mathbb{Z}_2$ Toric code.
More specifically, the equivalence is found by moving between the current lattice configuration and an edge lattice hosting the type $2$ qubits on the horizontal edges, and type $1$ qubits on the vertical edges~\cite{chen2023equivalence}.
This demonstrates that solutions to our plaquette constraints can be produced using unitary quantum circuits of depth $\order{L}$ (with $L$ being the size of the squared lattice), or using measurement-based feedback control in $\order{1}$ as explicitly presented in~\cite{Tantivasadakarn2024} for the Wen's plaquette model.

Although this derivation has been carried out only for the spinless mapping, it suggests that a similar generalization is possible for the spinful case, particularly for the ``spin-split mapping'', which is a direct extension of the spinless construction.

\section{Conclusions}
In this paper, we demonstrated how qudits, specifically ququarts, can be used to simulate local fermionic lattice Hamiltonians with local qudit operators. 
We introduced a mapping suitable for spinless fermions and two distinct mappings for spinful fermions.
These mappings were validated through simulations of quench quantum dynamics in the $t$-$V$ and Fermi-Hubbard model.
Each fermionic mode requires one ququart on the square lattice, in contrast to fermion-to-qubit mappings that require additional auxiliary qubits. This advantage can be exploited to reduce the operator weights and number of two-qudit gates in the Trotterized time evolution of lattice fermions.
Hence, our work offers a recipe for using the increase in the local Hilbert dimension of qudits to simulate local fermionic operators, without invoking additional (auxiliary) qudits. Hereby, we store additional parity information in the increased local Hilbert space.
Our results show how various associations of the fermionic modes with the qudit operators yield different qudit weight operators and circuit depths that can be optimized to improve the two-qudit gate requirements. Due to the improved scaling behavior and reduced two-qudit gates of our mappings, they can be used to simulate 2D materials with significantly larger system sizes in the future.

Future directions for this work include investigating gate compilation to hardware-specific gate sets, exploring other fermion-to-qudit identifications, and implementing compact encodings~\cite{derby2021compact} for qudit systems.
Another interesting study is the development of quantum-error-correction strategies tailored to fermionic simulations on qudits~\cite{jiang2019majorana}.
Generalizations to higher spatial dimensions and other geometries are conceptually straightforward within our framework, through the assignment of a qudit level to each edge attachment per node. Furthermore, using higher-dimensional qudits could enable simulation of systems with richer internal symmetries, such as $\mathrm{SU}(N)$ fermions or high-energy models.

\section{Acknowledgements}
This work is supported by the PNRR MUR projects PE0000023-NQSTI and CN00000013-ICSC and by QUART\&T, a project funded by the Italian Institute of Nuclear Physics (INFN) within the Technological and Interdisciplinary Research Commission (CSN5) and Theoretical Physics Commission (CSN4).
A.G. acknowledges support by the Horizon 2020 Marie Sklodowska-Curie action (H2020-MSCA-IF GA No.101027746). 

\section*{Data availability}
The data used in the simulations presented in this article, as well as the code employed to produce them, are openly available at~\cite{github}.

\bibliographystyle{quantum}
\bibliography{biblio}

\newpage
\onecolumngrid
\appendix

\section{Euclidean Dirac matrices\label{app:gamma_matrices}}
Before reporting the transformation matrices used, let us first introduce the $\Gamma$ matrices used throughout the paper. These matrices, which are one of the possible sets of operators to extend the Pauli operators to the ququart case, were introduced in Ref.~\cite{Vezvaee2024}, but we repeat them here for completeness
\begin{align*}
    \Gamma_1 = \sigma_x \otimes \iden_{2\times 2} = \left[\begin{smallmatrix}
        0 & 0 & 1 & 0 \\
        0 & 0 & 0 & 1 \\
        1 & 0 & 0 & 0 \\
        0 & 1 & 0 & 0
    \end{smallmatrix}\right]
    ,\quad &
    \Gamma_2 = \sigma_y \otimes \iden_{2
    \times 2} = \left[\begin{smallmatrix}
        0 & 0 & -i & 0 \\
        0 & 0 & 0 & -i \\
        i & 0 & 0 & 0 \\
        0 & i & 0 & 0
    \end{smallmatrix}\right],\\
    \Gamma_3 = \sigma_z \otimes \sigma_x = \left[\begin{smallmatrix}
        0 & 1 & 0 & 0 \\
        1 & 0 & 0 & 0 \\
        0 & 0 & 0 & -1 \\
        0 & 0 & -1 & 0
    \end{smallmatrix}\right]
    ,\quad &
    \Gamma_4 = \sigma_z \otimes \sigma_y = \left[\begin{smallmatrix}
        0 & -i & 0 & 0 \\
        i & 0 & 0 & 0 \\
        0 & 0 & 0 & i \\
        0 & 0 & -i & 0
    \end{smallmatrix}\right],\\
    \tilde\Gamma = \sigma_z \otimes \sigma_z = - \Gamma_1\Gamma_2\Gamma_3\Gamma_4  = \left[\begin{smallmatrix}
        1 & 0 & 0 & 0 \\
        0 & -1 & 0 & 0 \\
        0 & 0 & -1 & 0 \\
        0 & 0 & 0 & 1
    \end{smallmatrix}\right]
\end{align*}

\section{Unitary conjugation}\label{app:unitary_conj}

In this section, we detail the procedure for finding the unitary transformations used in \cref{sec:gate_comparison}.

In general, we wanted to find the solution to an equation of the form: 
\begin{equation}\label{eq:unitary_conj}
    e^{A\otimes B} = U^\dagger e^{D \otimes B} U
\end{equation}
where $A$ and $U$ are operators involving products of operators on different qudits, $A$ is hermitian, and $D$ is diagonal and therefore involves only separate single-qudit operations. The operator $B$, that acts on other qudits, is left unaltered by the transformation.

To determine the matrix $U$, we first compare the eigenvalues of matrices $A$ and $D$. If the eigenvalues are identical, we can select the eigenvectors of both operators and perform a QR decomposition to ensure orthonormality. The transformation matrix is then obtained as $U=\text{eigenvec}_B \text{eigenvec}_A^\dagger$.
This method guarantees that $U$ is unitary and satisfies \cref{eq:unitary_conj}.

Let's now consider the operator for the horizontal hopping terms in the second spinful mapping.
We define $A=i \left( \tilde\Gamma^1_{\vec r}\Gamma^2_{\vec{r+x}} -\tilde\Gamma^2_{\vec r}\Gamma^1_{\vec{r+x}}\right)$ and $D=\tilde\Gamma_{\vec{r}}+ \tilde\Gamma_{\vec{r+x}}$. Note that the choice of $D$ is not univocal. Applying the method just presented, we find that:
\begin{equation*}
U=
\left(\begin{smallmatrix}
0 & 0 & -\sqrt{0.5} & 0 & 0 & 0 & 0 & 0 & \sqrt{0.5} & 0 & 0 & 0 & 0 & 0 & 0 & 0\\
1 & 0 & 0 & 0 & 0 & 0 & 0 & 0 & 0 & 0 & 0 & 0 & 0 & 0 & 0 & 0\\
0 & 1 & 0 & 0 & 0 & 0 & 0 & 0 & 0 & 0 & 0 & 0 & 0 & 0 & 0 & 0\\
0 & 0 & 0 & \sqrt{0.5} & 0 & 0 & 0 & 0 & 0 & -\sqrt{0.5} & 0 & 0 & 0 & 0 & 0 & 0\\
0 & 0 & 0 & 0 & -1 & 0 & 0 & 0 & 0 & 0 & 0 & 0 & 0 & 0 & 0 & 0\\
0 & 0 & \sqrt{0.5} & 0 & 0 & 0 & 0 & 0 & \sqrt{0.5} & 0 & 0 & 0 & 0 & 0 & 0 & 0\\
0 & 0 & 0 & \sqrt{0.5} & 0 & 0 & 0 & 0 & 0 & \sqrt{0.5} & 0 & 0 & 0 & 0 & 0 & 0\\
0 & 0 & 0 & 0 & 0 & -1 & 0 & 0 & 0 & 0 & 0 & 0 & 0 & 0 & 0 & 0\\
0 & 0 & 0 & 0 & 0 & 0 & 0 & 0 & 0 & 0 & -1 & 0 & 0 & 0 & 0 & 0\\
0 & 0 & 0 & 0 & 0 & 0 & -\sqrt{0.5} & 0 & 0 & 0 & 0 & 0 & \sqrt{0.5} & 0 & 0 & 0\\
0 & 0 & 0 & 0 & 0 & 0 & 0 & -\sqrt{0.5} & 0 & 0 & 0 & 0 & 0 & \sqrt{0.5} & 0 & 0\\
0 & 0 & 0 & 0 & 0 & 0 & 0 & 0 & 0 & 0 & 0 & 1 & 0 & 0 & 0 & 0\\
0 & 0 & 0 & 0 & 0 & 0 & -\sqrt{0.5} & 0 & 0 & 0 & 0 & 0 & -\sqrt{0.5} & 0 & 0 & 0\\
0 & 0 & 0 & 0 & 0 & 0 & 0 & 0 & 0 & 0 & 0 & 0 & 0 & 0 & 1 & 0\\
0 & 0 & 0 & 0 & 0 & 0 & 0 & 0 & 0 & 0 & 0 & 0 & 0 & 0 & 0 & 1\\
0 & 0 & 0 & 0 & 0 & 0 & 0 & -\sqrt{0.5} & 0 & 0 & 0 & 0 & 0 & -\sqrt{0.5} & 0 & 0\\
\end{smallmatrix}\right)
\end{equation*}

For the vertical operators we need two different operators.
First we have $A=\tilde\Gamma^1_{\vec{r}}\tilde\Gamma^2_{\vec{r+y}} - \tilde\Gamma^2_{\vec{r}}\tilde\Gamma^1_{\vec{r+y}}$ and $D=\tilde\Gamma_{\vec{r}}+ \tilde\Gamma_{\vec{r+y}}$ and we find
\begin{equation*}
    V=\left(\begin{smallmatrix}
0 & 0 & i\sqrt{0.5} & 0 & 0 & 0 & 0 & 0 & \sqrt{0.5} & 0 & 0 & 0 & 0 & 0 & 0 & 0\\
1 & 0 & 0 & 0 & 0 & 0 & 0 & 0 & 0 & 0 & 0 & 0 & 0 & 0 & 0 & 0\\
0 & 1 & 0 & 0 & 0 & 0 & 0 & 0 & 0 & 0 & 0 & 0 & 0 & 0 & 0 & 0\\
0 & 0 & 0 & i\sqrt{0.5} & 0 & 0 & 0 & 0 & 0 & -\sqrt{0.5} & 0 & 0 & 0 & 0 & 0 & 0\\
0 & 0 & 0 & 0 & 1 & 0 & 0 & 0 & 0 & 0 & 0 & 0 & 0 & 0 & 0 & 0\\
0 & 0 & -i\sqrt{0.5} & 0 & 0 & 0 & 0 & 0 & \sqrt{0.5} & 0 & 0 & 0 & 0 & 0 & 0 & 0\\
0 & 0 & 0 & i\sqrt{0.5} & 0 & 0 & 0 & 0 & 0 & \sqrt{0.5} & 0 & 0 & 0 & 0 & 0 & 0\\
0 & 0 & 0 & 0 & 0 & 1 & 0 & 0 & 0 & 0 & 0 & 0 & 0 & 0 & 0 & 0\\
0 & 0 & 0 & 0 & 0 & 0 & 0 & 0 & 0 & 0 & -1 & 0 & 0 & 0 & 0 & 0\\
0 & 0 & 0 & 0 & 0 & 0 & i\sqrt{0.5} & 0 & 0 & 0 & 0 & 0 & \sqrt{0.5} & 0 & 0 & 0\\
0 & 0 & 0 & 0 & 0 & 0 & 0 & -i\sqrt{0.5} & 0 & 0 & 0 & 0 & 0 & \sqrt{0.5} & 0 & 0\\
0 & 0 & 0 & 0 & 0 & 0 & 0 & 0 & 0 & 0 & 0 & 1 & 0 & 0 & 0 & 0\\
0 & 0 & 0 & 0 & 0 & 0 & -i\sqrt{0.5} & 0 & 0 & 0 & 0 & 0 & \sqrt{0.5} & 0 & 0 & 0\\
0 & 0 & 0 & 0 & 0 & 0 & 0 & 0 & 0 & 0 & 0 & 0 & 0 & 0 & 1 & 0\\
0 & 0 & 0 & 0 & 0 & 0 & 0 & 0 & 0 & 0 & 0 & 0 & 0 & 0 & 0 & 1\\
0 & 0 & 0 & 0 & 0 & 0 & 0 & i\sqrt{0.5} & 0 & 0 & 0 & 0 & 0 & \sqrt{0.5} & 0 & 0\\
\end{smallmatrix}\right)
\end{equation*}

And finally we have $A=\Gamma^1_{\vec{r'}}\Gamma^2_{\vec{r'+y}}$ and $D=\tilde\Gamma_{\vec{r'}}$ that leads to:
\begin{equation*}
    W=\left(\begin{smallmatrix}
-i\sqrt{0.5} & 0 & 0 & 0 & 0 & 0 & 0 & 0 & 0 & 0 & -\sqrt{0.5} & 0 & 0 & 0 & 0 & 0\\
0 & 0 & -i\sqrt{0.5} & 0 & 0 & 0 & 0 & 0 & \sqrt{0.5} & 0 & 0 & 0 & 0 & 0 & 0 & 0\\
0 & 0 & 0 & 0 & -i\sqrt{0.5} & 0 & 0 & 0 & 0 & 0 & 0 & 0 & 0 & 0 & -\sqrt{0.5} & 0\\
0 & 0 & 0 & 0 & 0 & 0 & -i\sqrt{0.5} & 0 & 0 & 0 & 0 & 0 & \sqrt{0.5} & 0 & 0 & 0\\
\sqrt{0.5} & 0 & 0 & 0 & 0 & 0 & 0 & 0 & 0 & 0 & i\sqrt{0.5} & 0 & 0 & 0 & 0 & 0\\
0 & 0 & \sqrt{0.5} & 0 & 0 & 0 & 0 & 0 & -i\sqrt{0.5} & 0 & 0 & 0 & 0 & 0 & 0 & 0\\
0 & 0 & 0 & 0 & \sqrt{0.5} & 0 & 0 & 0 & 0 & 0 & 0 & 0 & 0 & 0 & i\sqrt{0.5} & 0\\
0 & 0 & 0 & 0 & 0 & 0 & \sqrt{0.5} & 0 & 0 & 0 & 0 & 0 & -i\sqrt{0.5} & 0 & 0 & 0\\
0 & 0 & 0 & \sqrt{0.5} & 0 & 0 & 0 & 0 & 0 & -i\sqrt{0.5} & 0 & 0 & 0 & 0 & 0 & 0\\
0 & \sqrt{0.5} & 0 & 0 & 0 & 0 & 0 & 0 & 0 & 0 & 0 & i\sqrt{0.5} & 0 & 0 & 0 & 0\\
0 & 0 & 0 & 0 & 0 & 0 & 0 & \sqrt{0.5} & 0 & 0 & 0 & 0 & 0 & -i\sqrt{0.5} & 0 & 0\\
0 & 0 & 0 & 0 & 0 & \sqrt{0.5} & 0 & 0 & 0 & 0 & 0 & 0 & 0 & 0 & 0 & i\sqrt{0.5}\\
0 & 0 & 0 & i\sqrt{0.5} & 0 & 0 & 0 & 0 & 0 & -\sqrt{0.5} & 0 & 0 & 0 & 0 & 0 & 0\\
0 & -i\sqrt{0.5} & 0 & 0 & 0 & 0 & 0 & 0 & 0 & 0 & 0 & -\sqrt{0.5} & 0 & 0 & 0 & 0\\
0 & 0 & 0 & 0 & 0 & 0 & 0 & i\sqrt{0.5} & 0 & 0 & 0 & 0 & 0 & -\sqrt{0.5} & 0 & 0\\
0 & 0 & 0 & 0 & 0 & i\sqrt{0.5} & 0 & 0 & 0 & 0 & 0 & 0 & 0 & 0 & 0 & \sqrt{0.5}\\
\end{smallmatrix}\right)
\end{equation*}

\section{Periodic boundary conditions}~\label{sec:pbc}

In the paper, we analyzed lattices with open boundary conditions. To extend our results to the periodic boundary conditions case, we need to consider an additional set of constraints.
In particular, other than elementary plaquettes, we will also have non-contractable Polyakov loops to be treated similarly to Eq.~\eqref{eq:identity_prod_A} 
\begin{align}
    \iden \eqconstr -i^{L_x} \prod_{m=1}^{L_x} A_{\vec{r}+(m-1)\vec{x}, \vec{r}+m\vec{x}} \\
    \iden \eqconstr -i^{L_y}\prod_{m=1}^{L_y} A_{\vec{r}+(m-1)\vec{y}, \vec{r}+m\vec{y}} 
\end{align}

\section{Jordan-Wigner Transformation for qudits}\label{app:jwt}

For completeness, here we report the Jordan Wigner Transformation (JWT) for fermion-to-qudit mapping, which is compared to the local mappings we introduced in this work in \cref{tab:gate_count_table}.
Following \cite{Vezvaee2024}, we use the $\Gamma$ matrices reported in the main text, defining the Clifford algebra $\mathcal C l_{0,4}$, and obeying the anti-commutation relations $\{ \Gamma^{\mu},\Gamma^{\nu}\} = 2\delta^{\mu,\nu}$.

JWT is performed by enumerating the degrees of freedom of the system (in the case of the lattice models presented in this work, the lattice sites) and then mapping each of them to a single ququart.
In this way, each ququart represents both the orbitals (spin up and spin down) of the lattice site, with no need for auxiliary ququarts.

\begin{align}
    \label{eq:jwt_qudit}
    \begin{split}
        f^{\dagger}_{m,\uparrow} & = \frac{1}{2} \prod_{k=0}^{m-1}\tilde{\Gamma}_{k} \cdot \left( \Gamma_{m}^{1} + i \Gamma_{m}^{2} \right)\\
        f_{m,\uparrow} & = \frac{1}{2}\prod_{k=0}^{m-1}\tilde{\Gamma}_{k} \cdot \left( \Gamma_{m}^{1} - i \Gamma_{m}^{2} \right)\\
        f^{\dagger}_{m,\downarrow} & = \frac{1}{2} \prod_{k=0}^{m-1}\tilde{\Gamma}_{k} \cdot \left( \Gamma_{m}^{3} + i \Gamma_{m}^{4} \right)\\
        f_{m,\downarrow} & = \frac{1}{2} \prod_{k=0}^{m-1}\tilde{\Gamma}_{k} \cdot \left( \Gamma_{m}^{3} - i \Gamma_{m}^{4} \right) \, ,
    \end{split}
\end{align}

where we indicated with a generic $m$ the site after the ordering and numbering.
As we can see in \cref{eq:jwt_qudit}, the main drawback is that we are introducing a non-local string of operators $\prod_{k=0}^{m-1}\tilde{\Gamma}_{k}$.
The non-local strings cancel each other in the one-dimensional Fermi-Hubbard model (as an example, the hopping term $f^{\dagger}_{m,\uparrow}f_{m+1,\uparrow} = \frac{1}{4}\left[ \left(\Gamma^{1}_{m} + i\Gamma^{2}_{m} \right)\tilde{\Gamma}_{m} \right] \Big[ \Gamma^{1}_{m+1} - i\Gamma^{2}_{m+1}\Big]$) but as soon as we address higher-dimensional systems these strings will not cancel and scale according to their dimensions.

\section{Exact Bosonization}\label{app:exact_bosonization}

\begin{figure*}
    \centering
    \includegraphics[width=0.8\linewidth]{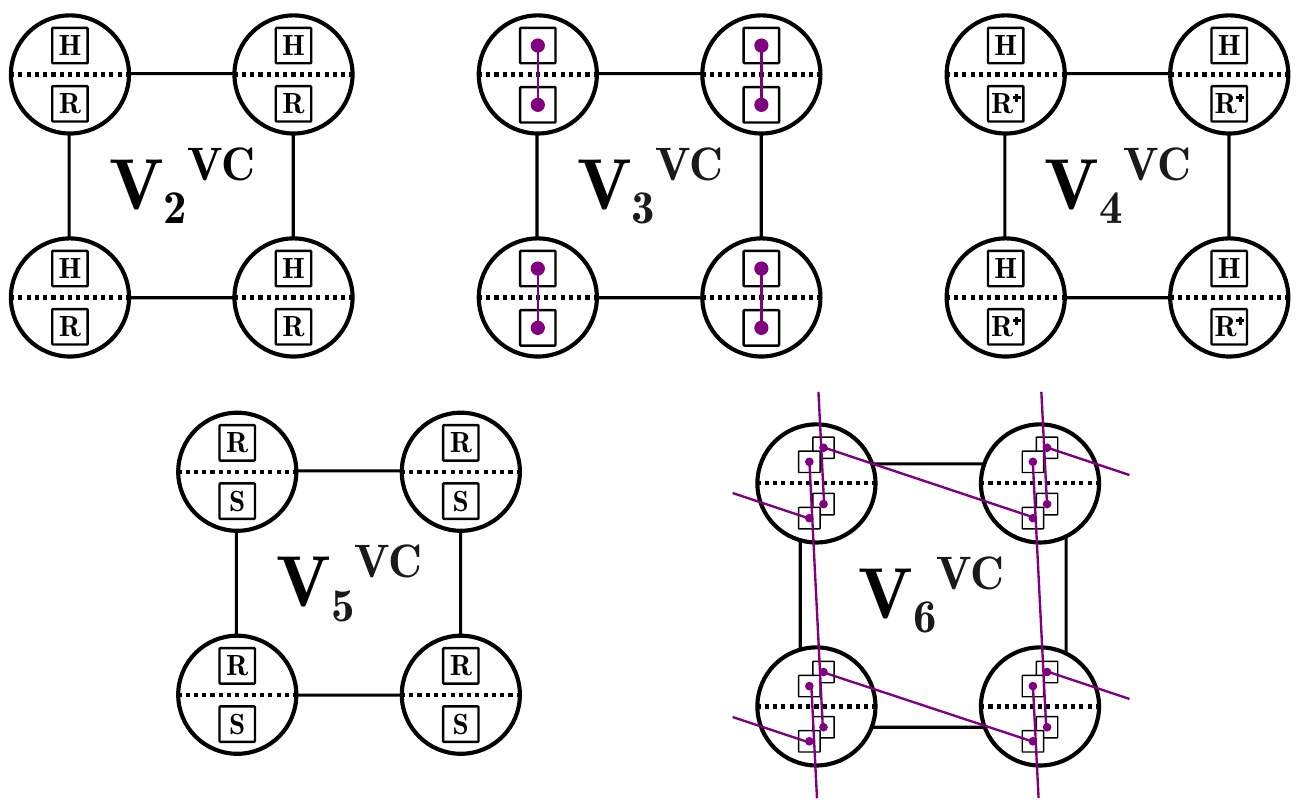}
    \caption{Graphical representation of the finite-depth circuits used to convert constraint $G_r$ to the exact bosonization. The total operator is defined as $V^{VC} = V_6^{VC} V_5^{VC} V_4^{VC} V_3^{VC} V_2^{VC}$.
    Figure adapted from \cite{chen2023equivalence}.}
    \label{fig:V_Vc}
\end{figure*}

In \cref{sec:toric_code}, we demonstrated the equivalence between the gauge constraint first introduced in \cref{eq:spinless_const_plaquette} and the operator of the Wen plaquette model. In Ref.~\cite{chen2023equivalence}, a quantum circuit is explicitly constructed to transform this operator into the gauge constraint of the exact bosonization.

In this appendix, we reproduce the same transformation, using circuits composed of the same gates, but focusing on their application to the ququart case, via the circuit shown in Fig. 13 of \cite{chen2023equivalence}.

We can identify the type-1 qubits as the ``top" two levels of each ququart, and the type-2 qubits as the ``bottom" two levels. The gauge constraint of exact bosonization, defined in Eq. (17) of \cite{chen2023equivalence}, is then obtained by unitary conjugation:
\begin{equation}
    G_{eb} = V^{VC} G_r (V^{VC})^\dagger
\end{equation}
where the unitary $V^{VC}$ is defined as a sequence of sub-circuits:
\begin{equation}
    V^{VC} = V_6^{VC} V_5^{VC} V_4^{VC} V_3^{VC} V_2^{VC}
\end{equation}
with each $V_i^{VC}$ defined in \cref{fig:V_Vc} in terms of qubit gates, explicitly given below, acting on subspaces of the ququarts.

The circuit $V_1^{VC}$ from \cite{chen2023equivalence} is omitted here, as it consists only of CNOT gates within each node between auxiliary and physical lattice. Since in our construction CNOT gates have already been applied to map $G_r$ to $G_v$ (in \cref{sec:toric_code}), applying $V_1^{VC}$ results in redundant CNOT$^2 = \mathbb{I}$ operations on each node.

Interestingly, all the $V_i^{VC}$ circuits consist of single-qudit gates, except for $V_6^{VC}$, which contains two-qudit entangling gates. In this latter case, care must be taken when comparing the circuit in \cref{fig:V_Vc} to that in Fig. 13 of \cite{chen2023equivalence}. Specifically, the horizontal and vertical axes must be swapped in order for our operator $G_v$ to match, graphically, the one presented Eq. (42) of \cite{chen2023equivalence}.

The qubit gates shown in \cref{fig:V_Vc} are defined as:
\begin{equation*}
    \boxed{H} = \frac{1}{\sqrt{2}}
    \begin{pmatrix}
        1 & 1 \\
        1 & -1
    \end{pmatrix}
    ,\quad 
    \boxed{S} =
    \begin{pmatrix}
        1 & 0 \\
        0 & i
    \end{pmatrix}
    ,\quad 
    \boxed{R} = \frac{1}{\sqrt{2}}
    \begin{pmatrix}
        1 & i \\
        i & 1
    \end{pmatrix}
    ,\quad
    \boxed{R^\dagger} = -\frac{1}{\sqrt{2}}
    \begin{pmatrix}
        1 & i \\
        i & 1
    \end{pmatrix}
    ,\quad 
    CZ =
    \begin{pmatrix}
    1 & 0 & 0 & 0 \\
    0 & 1 & 0 & 0 \\
    0 & 0 & 1 & 0 \\
    0 & 0 & 0 & -1 \\
    \end{pmatrix}
\end{equation*}

\end{document}